\documentclass[a4paper,fleqn,usenatbib,useAMS]{mnras}
\pdfoutput=1
\usepackage{amsmath,amssymb}
\usepackage{graphicx}	

\newcommand{\ok}{}
\newcommand{\jn}{}
\newcommand{\mpo}{}

\newcommand{\jnf}{}
\newcommand{\mpof}{}
\newcommand{\okr}{}
\newcommand{\jnr}{}
\newcommand{\mpor}{}

\usepackage{natbib}
\usepackage{comment}
\usepackage{color}

\title[Spatial evolution of Bell's instability]{Spatio-temporal evolution of the nonresonant instability in shock precursors of young supernova remnants}

\author[Kobzar, Niemiec, Pohl \& Bohdan]{
Oleh Kobzar,$^{1}$\thanks{E-mail: Oleh.Kobzar@ifj.edu.pl}
Jacek Niemiec,$^{1}$ 
Martin Pohl,$^{2,3}$
Artem Bohdan $^{1}$
\\
$^{1}$Instytut Fizyki J\c{a}drowej PAN, ul. Radzikowskiego 152, 31-342 Krak\'{o}w, Poland\\
$^{2}$Institute of Physics and Astronomy, University of Potsdam, 14476 Potsdam, Germany\\
$^{3}$DESY, 15738 Zeuthen, Germany
}

\date{Accepted XXX. Received YYY; in original form ZZZ}

\pubyear{2017}

\begin{document}
\label{firstpage}
\pagerange{\pageref{firstpage}--\pageref{lastpage}}
\maketitle

\begin{abstract}
A nonresonant cosmic-ray-current-driven instability \mpo{may operate} 
in the \ok{shock} precursors of young supernova remnants \mpo{and} be responsible for magnetic-field amplification, plasma heating \okr{and} turbulence. \mpo{Earlier simulations demonstrated magnetic-field amplification, and in \mpof{kinetic studies a reduction of the \jnr{relative} drift between cosmic rays and thermal plasma was observed as backreaction}. \okr{However,} all published simulations used periodic boundary conditions, 
which do not account for mass conservation in decelerating flows and \okr{only allow the temporal development to be studied}. Here we report results of \jnf{fully kinetic} Particle-In-Cell simulations with open boundaries that permit inflow of plasma on one side of the simulation box and outflow at the other end, hence allowing an investigation of both the temporal and the spatial development of the instability. 
Magnetic-field amplification proceeds as in \jnf{studies with periodic boundaries and, observed here for the first time, the \jnr{reduction of relative} \jnf{drifts} causes the formation of a shock-like compression structure at which a fraction of the plasma ions are reflected.}} 
Turbulent electric field \okr{generated by} the nonresonant instability inelastically scatters cosmic rays, modifying and  anisotropizing their energy distribution. \jnf{Spatial CR scattering is compatible with Bohm diffusion.}
\ok{Electromagnetic turbulence leads to significant nonadiabatic heating of the background plasma maintaining bulk equipartition between ions and electrons\okr{. The} highest temperatures are reached at sites of large-amplitude {\it electrostatic} fields. Ion spectra show supra-thermal tails resulting from stochastic scattering in the turbulent electric field. \mpor{Together, these modifications in the plasma flow will affect the properties of the shock and particle acceleration there.}}
\end{abstract}

\begin{keywords}
acceleration of particles -- cosmic rays -- shock waves -- ISM:supernova remnants -- methods:numerical -- turbulence
\end{keywords}



\section{Introduction}\label{introduction}

Galactic cosmic rays (CRs) with energies \mpo{up to a few peta-electronvolt} (PeV) are widely believed to originate from young shell-type supernova remnants (SNRs), at whose forward shock waves particles can be energized in the diffusive shock-acceleration (DSA) process \citep{2008ARA&A..46...89R}. The confinement of high-energy particles near the shock \mpo{and the inferred rate of shock crossings require magnetic turbulence amplified to levels much larger than those typically found} in the  interstellar medium (ISM).  
Observations of X-ray synchrotron emission from young SNRs, that is partially produced in narrow filaments \citep[e.g.,][]{2003ApJ...584..758V, 2005ApJ...621..793B} and can be variable on short time scales \citep{2007Natur.449..576U}, provide evidence for magnetic-field amplification by factors $10-100$ compared to the ISM level, \mpo{even if turbulence damping accounts for the filaments \citep{2005ApJ...626L.101P,2012A&A...545A..47R,2015ApJ...812..101T} and dynamic turbulence generates the variability \citep{2008ApJ...689L.133B}.} Detection of TeV-energy $\gamma$-ray emission from the same sources 
\citep[e.g.,][]{2011ApJ...730L..20A,2010ApJ...714..163A,2007A&A...464..235A,2009ApJ...692.1500A}
proves electron and/or proton acceleration to very high energies and suggests a direct relation between energetic particle production and non-linear magnetic-field amplification.
\jnr{The latter is also needed in systems hosting relativistic shocks, such as jets of radio galaxies \citep[e.g.,][]{2015ApJ...806..243A} or gamma-ray bursts \citep[GRB, see, e.g.,][]{2006ApJ...651..979M}.}

  
It has been long recognized on theoretical grounds that upstream of the shock magnetic fields can be amplified by the CRs themselves. Diffusing streaming of energetic particles in a quasi-parallel shock precursor can excite plasma instabilities generating magnetohydrodynamic (MHD) waves that are either resonant  \citep{1967ApJ...147..689L, 1969ApJ...156..445, wentzel74, 1975MNRAS.172..557S, 1975MNRAS.173..245S, 1975MNRAS.173..255S, 1983A&A...119..274A} or nonresonant with CRs. It was shown by \cite{bell04} that in young SNRs with efficient particle production the nonresonant CR streaming instability is the fastest one and generates short-scale magnetic turbulence with wavelengths much smaller than the gyroradii \mpo{of the CRs driving it}. The instability appears due to the current \mpof{density}, $\bmath{j}_{\rm CR}$, carried by CR protons drifting with respect to the magnetized upstream plasma. The most rapid growth 
occurs for a wavevector, $\bmath{k}$, aligned with \jnr{the mean homogeneous magnetic field} $\bmath{B}_0$ \citep{bell05}. In this case,
\jnr{and for $\bmath{j}_{\rm CR}$ parallel to $\bmath{B}_0$,}
the magnetic fluctuations take a form of a right-hand circularly polarized \mpo{wave with \jnr{the complex frequency $\omega$ and} maximum growth rate 
\begin{equation}
{\rm Im}\,\omega=\gamma_{\rm max}=v_{\rm A}k_{\parallel{\rm max}}
\end{equation}
at} the wavelength
\begin{equation}
\lambda_{\rm max}=\frac{4\pi B_{\parallel 0}}{\mu_0 j_{\parallel \rm CR}},
\end{equation}
where \jnr{the subscript `max' stands for the most unstable mode, $\mu_0$ is the permeability of free space and} $v_{\rm A}$ is the Alfv\'en velocity.
The CR current is balanced by the return current, $\bmath{j}_{\rm ret}=-\bmath{j}_{\rm CR}$, carried by the background plasma. The resulting Lorentz force, $\bmath{F}=\bmath{j}_{\rm ret}\times \bmath{B}$, acts on the plasma increasing the amplitude of the perturbed magnetic field \citep{zirakashvili08}. As the direction of the force is always perpendicular to the CR current, a purely growing (${\rm Re}\,\omega\approx 0$) mode results \okr{and} transverse magnetic-field components are amplified.
A plasma magnetization condition, $\omega\ll\Omega_i$, must apply, where $\Omega_i$ \mpof{is} the ion gyrofrequency. 
\mpo{The original MHD treatment of the instability was confirmed by kinetic studies \cite[e.g.][]{reville06,amato09}.}

\mpo{The nonlinear evolution of the nonresonant instability has been numerically} studied with MHD \citep{bell04, zirakashvili08, 2013MNRAS.435.1174S, 2014ApJ...788..107}, 
hybrid kinetic \citep{lucek, 2010ApJ...711L.127G} \okr{and} fully kinetic particle-in-cell (PIC) simulations \citep{niemiec_2008, 2009ApJ...694..626R, 2009ApJ...706...38S}.
\mpo{These studies naturally model only a small portion of the \jnr{far-upstream} shock precursor 
and all of them employed periodic boundary conditions, implying the spatial impact of the instability is ignored.
\mpor{The spatial structure of the \emph{perpendicular} case of the instability, 
in which $\bmath{j}_{\rm CR}\perp \bmath{B}_0$,  was studied by Riquelme \& Spitkovsky (2010). The primary effect of $\bmath{j}_{\rm CR}\perp \bmath{B}_0$ is a homogeneous linear acceleration of the upstream plasma with $\delta v= 2\,V_\mathrm{A}$ for each growth time of the instability. One may also envision localized regions with $\bmath{j}_{\rm CR}\perp \bmath{B}_0$ which would then be differentially accelerated, inducing vorticity and further magnetic-field amplification by turbulent dynamo action. This situation may arise for lower-energy CRs which are confined closer to the shock and can produce a transverse CR current by interacting with the turbulent field structures generated already far upstream by high-energy CRs via the nonresonant instability with $\bmath{j}_{\rm CR}\parallel\bmath{B}_0$, that is presumed to have provided large chunks of plasma with magnetic field perpendicular to the shock normal.}

The studies \jnr{using periodic boxes and field-aligned CR currents} demonstrated significant magnetic-field amplification by an initially parallel wave} with preferred wavelength $\lambda_{\rm max}$ as predicted by the linear theory. In the nonlinear phase, density fluctuations are imposed in the quasi-thermal plasma, and the plasma acquires strong transverse \mpo{velocity fluctuations}. The net $\bmath{j}\times \bmath{B}$ force pushes on the plasma that \mpo{eventually forms low-density cavities surrounded by filaments of compressed material}. As the cavities expand and merge with \mpo{each other}, the characteristic scale of magnetic-field fluctuations increases, but the turbulence growth eventually slows down or saturates. The exact saturation mechanism differs \mpo{with the choice of \jnf{the} simulation technique}, and the saturation levels vary considerably.  

MHD simulations use a fluid model for the background plasma and represent CRs as a constant external current, which means that backreaction on CRs is suppressed. The saturation is due to magnetic-field tension \mpof{opposing} the $\bmath{j}\times \bmath{B}$ force and scales with wavenumber \mpo{to yield a saturation level}
$B_{\rm sat}\sim \mu_0j_{\parallel \rm CR}/k$. Since the growth rate increases with \mpo{both the wavenumber and} the CR current, $\gamma\sim (kj_{\parallel \rm CR})^{1/2}$, the faster growing short-scale modes saturate earlier \mpo{and} at lower amplitudes than the slowly-growing longer-wavelength modes. \mpo{Consequently, as the turbulence evolves its characteristic wavelength also increases until it becomes commensurate with the CR Larmor radius, $k r_\mathrm{CRg}\approx 1$,} if the CR current is assumed constant. \jnr{However,} in reality the turbulence growth slows down once the magnetic-field amplitude increases to $\delta B/B_0\ga 1$ and density fluctuations appear, grow in size, acquire transverse motions \okr{and} start to \mpo{collide with} each other. \mpo{The rms field amplitude will still slowly grow, and a full saturation is never observed in MHD simulations \citep{zirakashvili08}, unless imposed by the boundary conditions and the finite size of the simulation box} \citep{bell04}. The final unsaturated field amplitudes obtained with \mpo{the MHD} approach reach $\delta B/B_0\simeq 30-50$.  

Fully kinetic PIC simulations allow treating both the electron and ion dynamics self-consistently and can thus account for the feedback \mpof{on CRs} of the amplified magnetic field. The backreaction causes a net transfer of momentum to the plasma. CRs slow down in bulk, and the background plasma is accelerated in the direction of the CR drift, reducing the relative speed and removing the CR current seen by the plasma.  
The resulting magnetic-field saturation level, $\delta B/B_0\simeq 10-20$, \mpo{is consistently lower than that obtained in MHD simulations} \citep{2009ApJ...694..626R,2009ApJ...706...38S}. 

Similar saturation mechanisms and levels were observed in PIC simulations of relativistic ion beams studied in the limit of magnetized ambient plasma \citep{niem10}, in which the turbulence \mpof{arises from growth} of the nonresonant modes \citep{reville06}. 
\jnr{Relativistic ion beams in this regime can be responsible for magnetic-field amplification upstream of external GRB shocks \citep{2006ApJ...651..979M} and at relativistic shocks associated with jets of active galactic nuclei. However, in such systems turbulence driven by the CR current}
can be accompanied with electromagnetic filamentation and electrostatic Buneman modes \jnr{\citep{niem10}}. The saturation physics is also consistent with quasi-linear predictions for nonresonant modes derived for nonrelativistic beams \citep{winske} and for CR streaming distributions with power-law momentum dependence \citep{luo}. Hybrid kinetic (fluid electrons and PIC model for ions) simulations by \citet{2010ApJ...711L.127G} yielded comparable amplification levels of $\delta B/B_0\simeq 10$ and noted the role of 
effective de-magnetization of the plasma through its rapid transverse motions. The importance of kinetic plasma treatment was further emphasized in PIC simulations that assumed a constant external CR current \mpof{as in MHD studies} \citep{ohira09,2009ApJ...694..626R}, \mpo{in which case the saturation is still caused by the reduction of the relative speed between CRs and plasma,} but now the plasma is accelerated to match the constant CR bulk drift \citep[see also][]{niem10}.    
  
\mpo{All previously published simulations used periodic boundary conditions, which suppress the spatial impact of the bulk acceleration of CRs and quasi-thermal plasma. In reality, continuity mandates compression or expansion to go along with bulk acceleration \citep{ohira09}, and CRs will continuously stream into fresh, unaffected plasma in which the instability must build up. The aim of this work is to study the nonlinear development and saturation processes of the nonresonant CR current-driven instability with open boundaries that permit inflow of plasma on one side of the simulation box and outflow at the other end. In this way the mass is conserved in the system \mpof{and} compression is naturally accounted for. \mpof{This} simulation represents the next step toward realistic modeling of nonresonant magnetic-field amplification in SNR shock precursors. Our study reproduces} some of the earlier results and also finds new effects specific to our new setup, that allows us to investigate both the temporal and the spatial development of the nonresonant instability. Particular attention is given to the microphysics of the saturation process, CR scattering \okr{and} plasma heating.

Although our approach offers a global model of the far-upstream \mpo{region of the} shock precursor, we do not directly include the shock that accelerates the CRs. Therefore, we cannot self-consistently determine the shock structure as modified by the CRs.      
Nonlinear steady-state shock models \mpo{\citep[e.g.,][]{zirakashvili08,2013MNRAS.435.1174S,2014ApJ...789..137B}} seek balance between magnetic-field amplification, that allows for CR confinement \mpof{and} enables rapid acceleration by the DSA process \okr{and} CR escape that provides the current driving the field generation
\citep{2013MNRAS.431..415B}. Our study can define conditions at which stationary solutions are possible and \mpof{aid in the validation of} the assumptions on particle scattering and heating which are important ingredients of these models.

The numerical PIC model is described in Section~\ref{setup} and results of our modeling are presented in Section~\ref{results}. Section~\ref{global} describes the main features of the nonresonant instability. The microphysics of the saturation process is discussed in Section~\ref{micro}, that treats CR scattering 
and plasma heating. 
\okr{We present a discussion in Section~\ref{discussion} and summarize our results in Section~\ref{summary}.}

\section{Simulation setup} \label{setup}

The nonresonant instability arises in the precursor of a strong nonrelativistic parallel shock
when shock-accelerated CRs drift against the cold upstream electron-ion plasma.
Our earlier simulations of this instability with periodic boundary conditions were conducted in the plasma-ion rest frame \jn{\citep{niemiec_2008,2009ApJ...706...38S}}. 
In this reference frame, an isotropic population of \mpof{relativistic} CR ions with \jnr{number} density $N_\mathrm{CR}\ll N_e$ moves with the shock velocity, $v_\mathrm{sh}$, along a homogeneous magnetic field. The current density carried by CR ions, 
$j_\mathrm{CR}=eN_\mathrm{CR}v_\mathrm{sh}$, \jnr{where $e$ is the electric charge}, is compensated by the return current density, $j_\mathrm{ret}=-eN_\mathrm{e}v_\mathrm{d}=-j_\mathrm{CR}$, provided by electrons of \jnr{number} density $N_\mathrm{e}$ that slowly drift relative to background ions with $v_\mathrm{d}=v_\mathrm{sh}N_\mathrm{CR}/N_\mathrm{e}$. \mpo{We add an} \emph{excess} electron population with \jnr{number} density $\delta N_\mathrm{e}=N_\mathrm{e}-N_\mathrm{i}=N_\mathrm{CR}$ to maintain charge neutrality. 

\begin{figure}
\centering
\includegraphics[width=\linewidth]{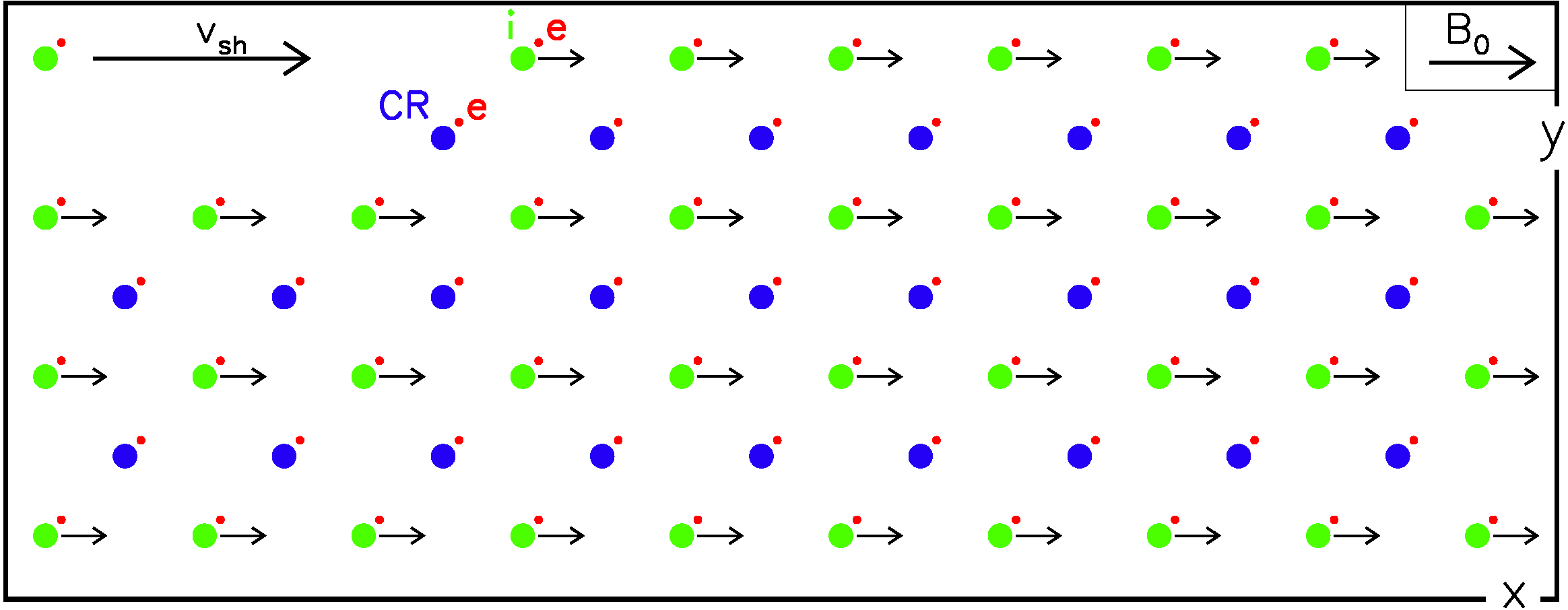}
\caption{\mpo{Sketch of the simulation setup. Cosmic rays ({\it blue}) and the associated excess electrons ({\it red}) permeate the entire simulation box. Pairs of plasma ions ({\it green}) and electrons ({\it red}) also fill the box, but move as ensemble with speed $v_\mathrm{sh}$ in the $x$-direction along the homogeneous magnetic field, \jn{$B_0=B_{\parallel 0}$}.} }
\label{sim-setup}
\end{figure}
\mpo{Maintaining stable conditions with open boundaries requires that the calculations in this work be} performed in the \emph{cosmic-ray rest frame}. \mpo{As illustrated in Fig.~\ref{sim-setup}, in}
the CR rest frame the electron-ion plasma forms a beam \mpo{with velocity $v_{sh}$. Charge compensation is established with \emph{excess} electrons that are initiated at rest to maintain current balance. With this method stable injection of the electron-ion beam into the numerical box is simpler to achieve than with, e.g., adding the \emph{excess} electrons to the electron-ion beam.}  

\mpo{In the linear phase the nonresonant mode has its peak growth rate at a wavelength smaller than the} \jn{CR gyroradius}, $\lambda_\mathrm{max} \ll r_\mathrm{CRg}$. \mpo{The simulation results presented here} are based on 
\jn{a PIC experiment utilizing}
an extremely large two-dimensional computational grid of size $L_x\times L_y= 130,000\times 12,000$ cells and a total of $5.6\cdot 10^{10}$ macro-particles. \mpo In terms of $\lambda_\mathrm{max}$, the size of the numerical box is $L_x\times L_y\simeq (293\times 27)\,\lambda_\mathrm{ max}^2$.
All system constituents -- CR ions, electrons and ions of the ambient plasma beam \okr{as well as} the electrons of the \emph{excess} population -- initially fill up the whole computational box.
Periodic boundary conditions for particles and electromagnetic fields are applied in the transverse ($y$) direction. The right and left box boundaries are open for the ambient beam particles and fields. 
Fresh particles of the electron-ion beam are injected at the \emph{left} side of the box, where \mpo{absorbing boundary conditions are applied} for fields. 
Outflowing beam particles are removed at the \emph{right} side of the box, 50 cells off the grid boundary. Here the \mpo{mean fluxes of electrons and ions} are nearly identical, and small-scale electromagnetic impulses generated by local charge imbalances propagate freely towards the open \jn{\it{right}} boundary for fields, at which they are effectively absorbed. 
CR ions are kept inside the box and reflected, if they encounter the open boundaries for the ambient plasma. 
The electron-ion plasma beam pushes the excess electrons to the right, leaving the much heavier CRs initially unaffected. The simulated system quickly adjusts itself to the new situation in a numerically stable way, producing a suitable return current and compensating for the small charge imbalance, exactly as one would envision a real plasma to respond. 
The excess electrons concentrate towards the right boundary for particles. They are not removed from the system in order to maintain charge conservation. 
On account of the high ambient plasma-to-CR density ratio, the concentrations in excess electrons are always a small perturbation to the total electron distribution, even in the nonlinear stage of the instability, during which density and field structures are produced that focus the plasma into compact islands. Therefore, the excess electrons do not play a significant role in the dynamics of the system.     
Numerical artifacts at the right boundary cause field fluctuations with much smaller amplitude than \mpof{those} of the turbulence in the region of interest\okr{. These} fluctuations do not radiate into the \jn{box} in any significant form. 
Therefore, the plasma outflow boundary does not have a noticeable influence on the evolution of \jn{Bell's modes} in our numerical experiment.

The numerical \mpo{parameters are similar to those used in our earlier simulation with periodic boundaries} \citep{2009ApJ...706...38S}, to permit a \jn{direct} comparison.
\jnr{As in that study, we also choose $\bmath{j}_{\rm CR}$ to be aligned with $\bmath{B}_0$, in which case the turbulence takes on the simplest form \citep{bell05}. The chosen parameters were shown in \citet{2009ApJ...706...38S} to well}
\mpo{resolve} all physical characteristics of the nonresonant instability. 
\mpor{The key issue in properly reproducing the instability characteristics is to assure the so-called magnetization condition, $\gamma_{\rm max}/\Omega_i\ll 1$. This condition does not explicitly depend on specific values of $\gamma_{\rm CR}$ or $N_{\rm CR}/N_i$, but these quantities must be chosen so that the instability condition is fulfilled. It is thus justified to say that our simulation probes the far-upstream structure of the precursor with high-energy cosmic rays. The rationale for selecting specific parameters was extensively given in Niemiec et al. (2008) and in Stroman et al. (2009), and among the drivers is the low growth rate for high-energy CRs of realistic density, that does not permit PIC simulations even in 2D. This rationale is supported with numerical tests, in particular for the range of $N_{\rm CR}/N_i$ in Niemiec et al. (2008) (also in Niemiec et al., 2010) and $\gamma_{\rm CR}$ in Stroman et al. (2009).}
The electron skin depth 
 $\lambda_{\rm se}=c/\omega_{\rm pe}=4\Delta$, where $\Delta$ is the grid cell size, $c$ is the speed of light, $\omega_{\rm pe}=\sqrt{e^2N_\mathrm{e}/\epsilon_0m_\mathrm{e}}$ the electron plasma frequency \okr{and} $m_\mathrm{e}$ \okr{is} the electron mass. 
CR ions are represented by a population of isotropic and monoenergetic particles with Lorentz factor $\Gamma_{\rm CR}=50$. The \jnr{number} density ratio between plasma ions and CR ions is $N_\mathrm{i}/N_\mathrm{CR}=50$,
and the assumed plasma magnetization, given as the ratio of the peak growth rate to the nonrelativistic ion gyrofrequency, is
$\gamma_{\rm max}/\Omega_\mathrm{i}=0.4$. 
The \mpo{homogeneous magnetic field, $B_0=B_{\parallel 0}$, is chosen for} \jn{a CR gyroradius} $r_{\rm CRg}\simeq 140,000\Delta$ 
\jn{and an Alfv\'en speed $v_{\rm A}=0.01c$, with $v_{\rm A}=[B_{\parallel 0}^2/\mu_0 (N_\mathrm{e}m_\mathrm{e}+N_\mathrm{i}m_\mathrm{i}]^{1/2}$}. 
The ambient plasma beam \mpo{moves in the $x$-direction, parallel to the uniform magnetic field, with velocity $v_\mathrm{sh}=0.4c$. 
The initial thermal speed of electrons, $v_\mathrm{e,th}=0.01c$, is the same for both plasma and {\it excess} electrons}. 
\okr{The} ions are in thermal equilibrium with the electrons. 
Computational constraints force us to use a reduced ion-to-electron mass ratio, $m_\mathrm{i}=50\,m_\mathrm{e}$, \mpo{for which the spatial and temporal scales} are still clearly separated. The wavelength of the most unstable mode is $\lambda_{\rm max}\simeq 444.3\Delta$ \okr{and} the inverse growth rate \okr{is} $\gamma_{\rm max}^{-1}\simeq \ok{\textrm{14,142}}\,\delta t$, where the timestep $\delta t=0.125/\omega_{\rm pe}$.

We follow the system evolution through the strongly nonlinear phase up to $t\,\gamma_{\rm max}=26.9$. 
\jnr{Beyond this time the largest scales of the turbulence structure start to become comparable to the transverse size of our numerical box, $L_y$, and cannot thus be properly resolved.}
The simulation uses $N_{\rm ppc}=9$ particles per cell per species, and statistical weights, $w_g=0.02$, are applied to CRs and excess electrons to establish the desired density \jn{ratio}. Pairs of ions and electrons are initiated at the same locations, so that the initial charge and current vanish. 
To stabilize the plasma against nonphysical effects on very long time scales and with a moderate number of particles per cell
we apply the numerical PIC model used recently in \citet{niem12}. The model involves second-order particle shapes (TSC -- Triangular-Shaped Cloud) and uses a second-order FDTD field-solver with a weak Friedman filter
\citep{green04} to suppress small-scale noise.
The simulations were performed with a modified version of the TRISTAN code \citep{tristan} parallelized using MPI~\citep{niemiec_2008}. We follow all three components of the particle velocities and electromagnetic fields.
\jnr{Note that this 2D3V model was demonstrated to well reproduce the results of fully three-dimensional kinetic simulations, including both the magnetic field amplification level and the turbulence structure \citep{niemiec_2008,2009ApJ...694..626R}. For the unstable wavevector aligned with $\bmath{B}_0$ and contained in the simulation plane, spatial distributions of particle densities and electromagnetic fields are projections onto a 2D plane of the turbulent structures observed in three-dimensions with kinetic and also MHD simulations. The neglect of gradients in z-direction inherent to 2D3V simulations may somewhat impact the development of 3D turbulence structures in the deep non-linear phase of the instability, e.g. concerning the size and lifetime of plasma voids, but that is more than balanced by the large benefit of our 2D3V numerical experiment covering a much wider range of dynamical scales of nonlinearly evolving turbulence than can be possibly afforded with a 3D simulation under current computing capabilities.}

The numerical model used in this study has been extensively tested. Test simulations with periodic boundary conditions in all directions verified that the physical features of the nonresonant instability (the growth rate, the wavelength of the most unstable mode, the nonlinear response) in the cosmic-ray rest frame match exactly \mpo{those obtained in the plasma} rest frame with the same parameters \citep{2009ApJ...706...38S}. 
The results do not significantly change if the system resolution (the electron skindepth) or \mpo{the number of particles per cell,}
$N_{\rm ppc}$, are increased. The numerical stability of plasma injection and \mpo{escape at the open boundary} has been \mpof{verified} in test simulations with an electron-ion beam propagating in \mpo{an otherwise empty box. Stability at the left boundary is ensured by the charge-neutral injection of particles}. 
Radiation-absorbing boundary conditions are used as originally implemented in the TRISTAN code \citep{tristan}.

\begin{figure*}
\centering
\includegraphics[width=\textwidth]{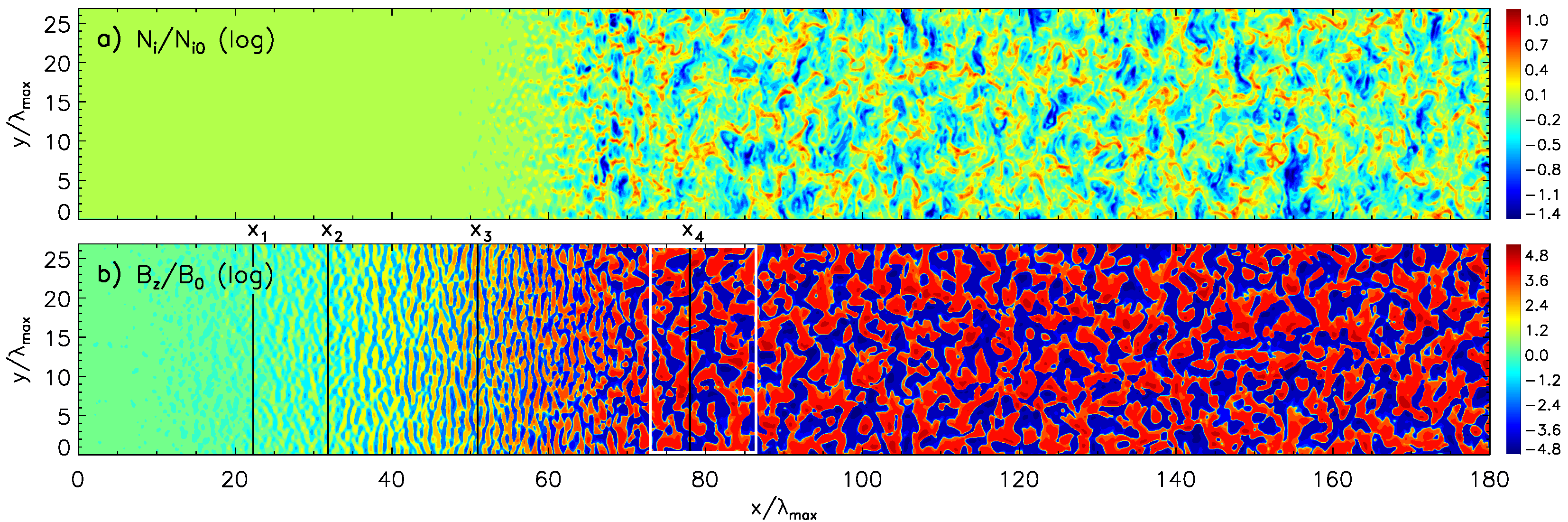}
\caption{Distribution of (a), the normalized ion \jnr{number} density, 
\okr{$(N_\mathrm{i}/N_{\rm i0})$}, 
and (b), the normalized magnetic-field component 
\okr{$(B_z/B_{\parallel 0})$} 
at time $t\gamma_{\rm max}= 13$, when the magnetic-field amplification is close to saturation. 
\okr{Logarithmic scaling is applied for density.} \jnr{The scaling for $B_z$ is also logarithmic, but sign-preserving, ($\mathrm{sgn}(B_z)\,(3+\log[\max(10^{-3},|B_z|/B_{\parallel 0})])$). 
The value $\pm 3$ hence corresponds to $B_z=\pm B_{\parallel 0}$, and field amplitudes below $10^{-3}\,B_{\parallel 0}$ are not resolved.}
Locations $(x_1, x_2, x_3, x_4)/\lambda_\mathrm{max} = (22.5, 32, 51, 75)$ marked with 
\okr{black} vertical lines indicate different phases of the system evolution discussed in the text and correspond to time instances
$t_1, t_2, t_3,$ and $t_4$ denoted in Fig.~\ref{avmag}. The white rectangular box in Fig.~\ref{bell1}b shows the region in which spatially averaged magnetic field amplitude shown with dashed lines in Fig.~\ref{avmag} is calculated at $t\gamma_{\rm max}=t_4=13$.
Only a portion of the simulation box is shown.} 
\label{bell1}
\end{figure*}

\begin{figure*}
\centering
\includegraphics[width=\textwidth]{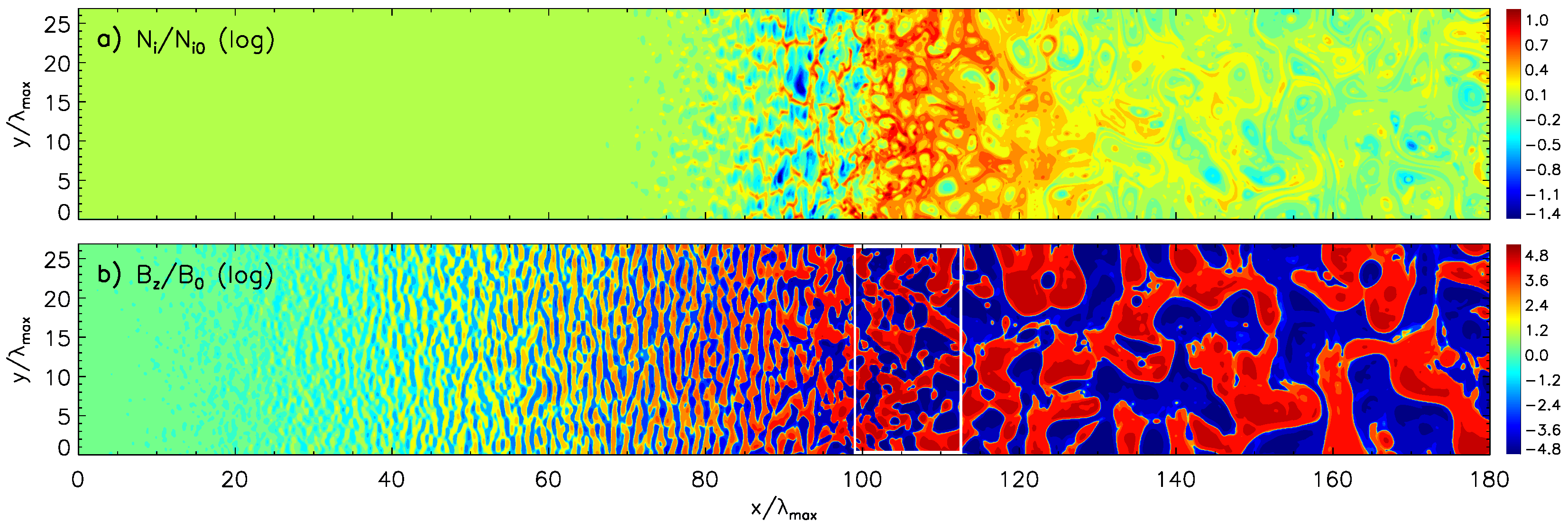}
\caption{\mpo{The distribution of ion \jnr{number} density and perpendicular magnetic field at the end of the numerical experiment at $t\gamma_{\rm max}= 26.9$. The display including} the dynamic range is the same as that used in Fig.~\ref{bell1}.} 
\label{bell2}
\end{figure*}

\section{Results} \label{results}

\mpo{The open boundaries used in this numerical study allow us to investigate both the temporal and the 
spatial development of Bell's nonresonant instability, and they permit maintaining mass conservation in the system.} In Section~\ref{global} we demonstrate magnetic-field amplification through the nonresonant instability, its saturation \okr{and} the main features of its backreaction on CRs that re-confirm our earlier results obtained in PIC simulations with periodic boundaries. We also describe new effects specific to our current, more realistic simulation setup. Section~\ref{micro} presents the microphysical picture of the nonlinear particle-wave interactions during the saturation phase that result in kinetic-energy transfer, CR scattering \okr{and} plasma heating. 

\subsection{\okr{Nonresonant instability: main features}\label{global}}

\mpo{In this section we describe the state of the system at time $t\gamma_{\rm max}= 13$, when the magnetic-field growth reaches saturation, and at time $t\gamma_{\rm max}= 26.9$, when the system is in the strongly nonlinear phase.} \jnr{We present a global spatial structure of the far-upstream precursor in Section~\ref{spatial}. Spatio-temporal development of magnetic field turbulence is described in Section~\ref{temporal}, the effects of the nonlinear plasma response in Section~\ref{plasma_turb} and the electric field turbulence in Section~\ref{efield}. Finally, in Section~\ref{phase-space} we demonstrate the effects of a compression wave -- a new feature revealed through our modeling with open boundaries -- on the field turbulence and particles in the precursor.}


\subsubsection{\jnr{Spatial structure of far-upstream precursor}\label{spatial}}

Figs.~\ref{bell1} and \ref{bell2} show the distributions of the transverse magnetic-field component, $B_z$, and the \jnr{number} density of ambient ions at the two points in time, respectively. 

To be noted from Figs.~\ref{bell1} and \ref{bell2} is the spatial imprint of the \mpo{time development of the instability that arises from the motion of the wave-carrying plasma from left to right. Growth of
magnetic turbulence becomes evident} at $x/\lambda_{\rm max}\approx 22$ and reaches its saturation at $x/\lambda_{\rm max}\approx 75$. 
In studies with periodic boundaries only the temporal evolution was observable. 
As \jn{in the present setup} the entire simulation box is filled with CRs and with streaming plasma at the beginning, 
\jn{the instability development is \mpof{mainly temporal in the right part} of the box in which the beam particles have travelled the same distance on the CR background from the time of their initial injection.}
Wave growth \jn{begins} only \jn{close to} the left boundary,
\jnr{where fresh plasma is injected,}
and as the instability needs a few growth times to reach a significant amplitude, the wave-carrying plasma will have streamed a certain distance. Later phases of turbulence development need more time, and hence the plasma will have propagated a larger distance. In the end, the temporal stages of the instability -- exponential growth, nonlinear response, saturation and cascading -- are mapped to certain locations in $x$ that slowly move to the right on account of the displacement of the CRs,
\jn{that are pushed on by the streaming plasma in the nonlinear phase}. Travel time is \jn{thus} unimportant 
\jn{for $x/\lambda_{\rm max}\gtrsim 75$ at $t\gamma_{\rm max}= 13$ (Fig.~\ref{bell1}) and for}
$x/\lambda_{\rm max}\gtrsim 100$ \jn{at the end of the simulation (Fig.~\ref{bell2}), where} we primarily observe the temporal development.

\jnr{One feature of nonlinear feedback \jn{in periodic simulations using \it{plasma-ion rest frame}} was bulk deceleration of cosmic rays and \jn{acceleration of} the thermal plasma until the remaining streaming was insufficient to further drive turbulence. As we discuss in Section~\ref{temporal}, the same effects are observed in our current study performed in the CR rest frame. One consequence of this backreaction in our realistic modeling with open boundaries is the appearance of a strong density enhancement, clearly visible in Fig.~\ref{bell2}a at $x/\lambda_{\rm max}\approx 100 - 130$. \mpor{This compression is caused by the collision of the incoming electron-ion plasma with the decelerated plasma in the turbulence zone. The compression factor still increases at the end of the simulation and} approaches the value of 4, as in a strong nonrelativistic shock. Detailed features of this shock-like structure will be described in the sections that follow.}


\subsubsection{\okr{Spatio-temporal \jnr{evolution of magnetic} turbulence}\label{temporal}}

\begin{figure}
\centering
\includegraphics[width=\linewidth]{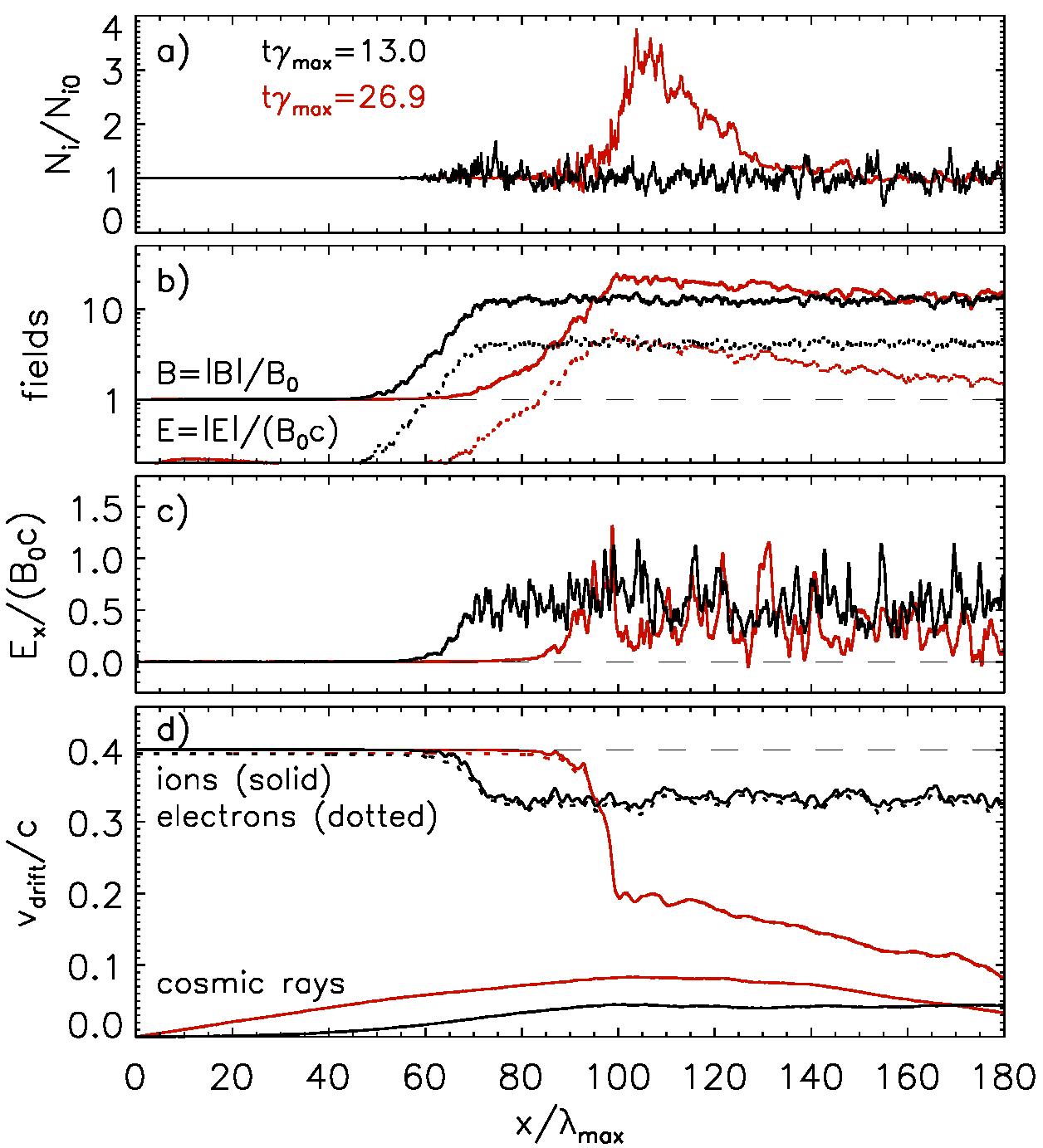}
\caption{\mpo{Spatial profiles of the ion \jnr{number} density (a), magnetic and electric field amplitudes (b), $E_x$ component of the electric field (c) \okr{and} particle velocities (d). \emph{Black} lines refer to time $t\gamma_{\rm max}= 13$, \okr{whereas} \emph{red} lines indicate the status at $t\gamma_{\rm max}=26.9$. In panel (b), the electric-field amplitude is displayed with \emph{dotted} lines. In panel (d), \emph{dotted} lines are for plasma electrons, and \emph{solid} lines beginning at $v/c=0.4$ indicate the \jnr{number} density of plasma ions, whereas \emph{solid} lines beginning at $v/c=0$ are for CR ions.} \jn{Profiles are calculated for a portion of the simulation box shown in Figs.~\ref{bell1}~and~\ref{bell2}}.}
\label{profiles}
\end{figure}

\mpo{The spatial and the temporal development \jnr{of the instability} are connected through the streaming of the plasma, and proper interpretation requires knowledge of its bulk velocity. We have averaged over coordinate $y$ various plasma variables to produce their spatial profiles in coordinate $x$. Fig.~\ref{profiles} displays such profiles of the ion \jnr{number} density, the magnetic and electric-field amplitudes, the $E_x$ component of the electric field \okr{and} particle velocities.}


\mpo{To be noted from Fig.~\ref{profiles}\okr{b} is the magnetic-field amplification to about $20\,B_{\parallel 0}$ at $t\gamma_{\rm max}= 26.9$, slightly exceeding that seen in simulations with periodic boundaries \citep{2009ApJ...706...38S} on account of compression. The bulk acceleration\jnr{/deceleration} observed \jnr{in earlier studies} is also evident \jnr{(Fig.~\ref{profiles}d)}, and at the location of the peak amplitude of the mean magnetic field the bulk velocities of plasma and CRs have converged to $\delta V\lesssim 0.05\,$c. 
Associated plasma compression by up to a factor 4, \jnr{identified already in Section~\ref{spatial},} is seen in \jn{Fig.~\ref{profiles}a} \jnr{for $x/\lambda_{\rm max}\approx 90-130$}. \jnr{E}ssentially all of it arises late in the evolution, \jnr{starting from $t\gamma_{\rm max}\approx 13.0$, close to the instability saturation time, and continuing until the end of the simulation at}
$t\gamma_{\rm max}= 26.9$. The onset of compression moves with the plasma flow that pushes on the turbulent zone, and at times $t\gamma_{\rm max}=13$ and $26.9$ it is contained in the white rectangular boxes in Figs.~\ref{bell1} and \ref{bell2}, respectively. 


\mpo{Knowledge of the plasma bulk speed permits decoupling of the temporal and the spatial evolution of the instability, \mpof{at least in the left part of the simulation box}. For that purpose we have defined 4 locations labeled $x_1$, $x_2$, $x_3$ \okr{and} $x_4$ and marked them in Fig.~\ref{bell1}. We also marked in white a region around $x_4$. These 4 locations correspond to the first emergence of waves (1), their exponential growth (2), entering the regime $\delta B \gtrsim B_{\parallel 0}$ (3) \okr{and} saturation \jn{(4)}. As wave-particle interactions can begin only at the left boundary of the simulation box \okr{and} a certain distance implies a specific streaming time, the four locations correspond to points $t_1$, $t_2$, $t_3$ \okr{and} $t_4$ in the temporal evolution of the instability. }

\mpo{Recalling that the streaming time for 
$135 \lesssim x/\lambda_{\rm max}\lesssim 180$ is not smaller than the simulation time, implying we primarily observe the temporal evolution, we can directly compare our new results with the earlier investigations. In Fig.~\ref{avmag} we plot \jnf{with {\it solid} line\jnr{s} the time evolution of the magnetic-field amplitude\jnr{s} spatially averaged in the region $135 \lesssim x/\lambda_{\rm max}\lesssim 180$, \jn{that does not contain the compression wave}. We add the time markers $t_1$ -- $t_4$ and can now compare the state of the system with that observed at the corresponding locations $x_1$ -- $x_4$ in Fig.~\ref{bell1}.}

\begin{figure}
\centering
\includegraphics[width=\linewidth]{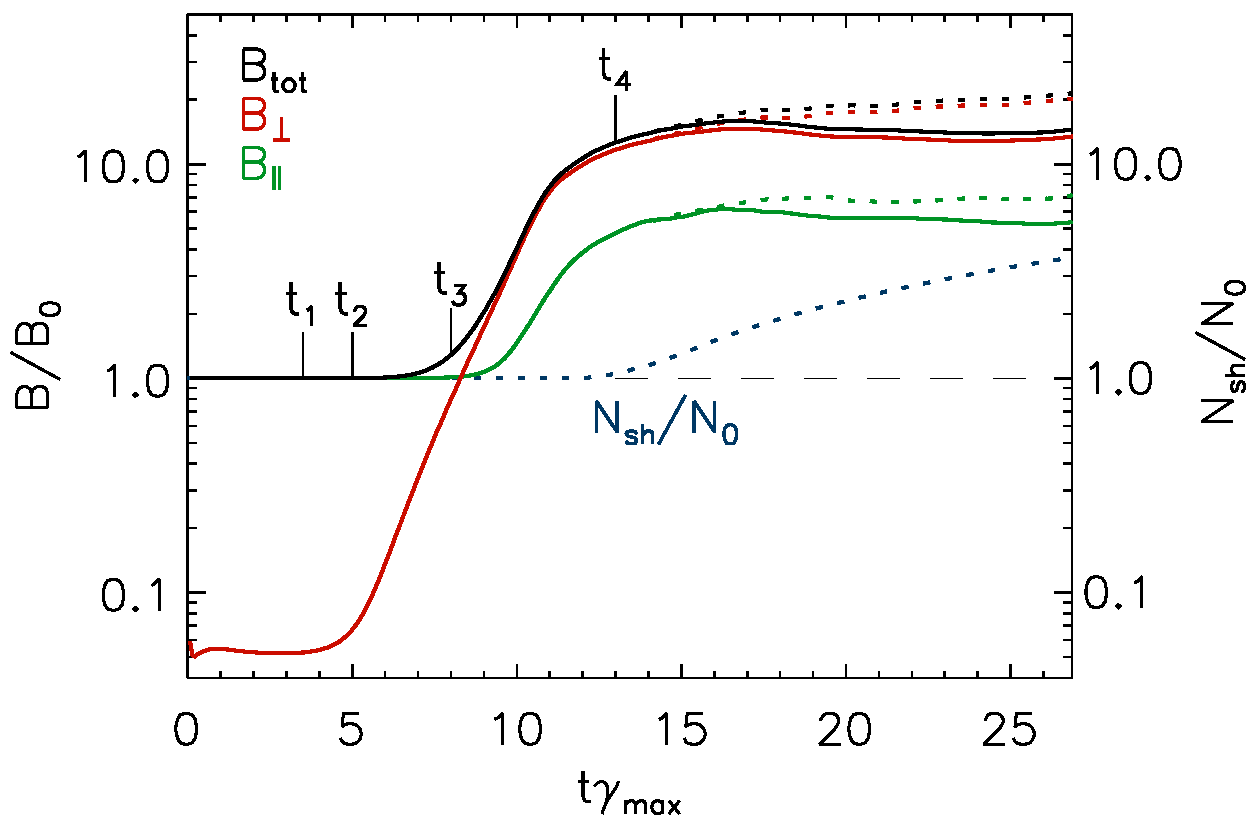}
\caption{Time evolution of the spatially averaged \mpo{and normalized magnetic-field strength \okr{and plasma density in the compression region}. \okr{Field} amplitudes averaged in the region between $x/\lambda_{\rm max}\approx 135$ and $x/\lambda_{\rm max}\approx 180$ are shown with \emph{solid} lines. \emph{Dashed} lines show the field amplitudes averaged in a region of thickness of $11\lambda_{\rm max}$ centered on and co-moving with the compression wave, that is marked with a white box in Figs.~\ref{bell1} and \ref{bell2}.} Magnetic-field components perpendicular and parallel to the initial homogeneous field, $B_0$, are marked with \emph{red} and \emph{green} lines, respectively. \emph{Black} lines denote the total magnetic-field strength. 
\mpor{The development of the compression region is indicated by the \emph{blue dashed} line that traces the density evolution in that region.}
Refer to Fig.~\ref{bell1} for explanation of time markers $t_1 - t_4$.} 
\label{avmag}
\end{figure}

The waves that start to emerge at $x/\lambda_{\rm max}=x_1\gtrsim 22.5$ (Fig.~\ref{bell1}) move with the beam (i.e., they are stationary in the ambient plasma rest frame) and have wave vectors parallel to the plasma drift with wavelength $\sim\lambda_{\rm max}$. \mpo{At this location the electron-ion beam has propagated through the CR plasma for $t\gamma_{\rm max}=t_1\approx 3.5$, and we see in Fig.~\ref{avmag} that at this time the first sign of growth in $B_\perp$ appears.
At $x/\lambda_{\rm max}=x_2\approx 32$ the waves are clearly visible in Fig.~\ref{bell1} \okr{and} in the temporal development depicted in Fig.~\ref{avmag} the exponential growth  with rate $\gtrsim 0.9\,\gamma_{\rm max}$ begins.
At $t\gamma_{\rm max}=t_3\approx 8$ the turbulent field} becomes comparable to the strength of the homogeneous magnetic field, 
$\langle\delta B/B_{\parallel 0}\rangle\ga 1$. \mpo{We observe at the corresponding location $x_3$ that the magnetic fluctuations start} to deviate from a purely parallel mode, and
the dominant length scale begins to exceed  $\lambda_{\rm max}$. The subsequent evolution leads to isotropic turbulence \mpo{beyond $x/\lambda_{\rm max}=x_4\approx 75$, corresponding to the saturation at $t\gamma_{\rm max}\gtrsim t_4=13$ at the level} $\langle\delta B/B_{\parallel 0}\rangle\approx 15$. At later times the mean field amplitude stays at roughly the same level but the turbulence scale quickly evolves to very long wavelengths, as is seen in Fig.~\ref{bell2} for $x/\lambda_{\rm max}\ga 110$.

The magnetic-field is amplified to essentially the same amplitude as
in our earlier simulations using periodic boundaries \citep{2009ApJ...706...38S}. \mpo{Fig.~\ref{avmag} also displays with \emph{dashed} lines the magnetic-field amplitudes in the compression region where, as noted before, larger amplification is reached.}
\jnr{One can note that magnetic-field amplification in the compression wave grows steadily, and by the end of our numerical experiment the field has an amplitude $\sim 30\%$ higher than that provided through the nonresonant instability \jn{(compare Fig.~\ref{profiles}b)}. It does not reach saturation, which suggest that the compression structure has not yet \mpof{attained a steady state}\okr{, as shown by the \emph{blue dashed} curve in Fig.~\ref{avmag}.}
\jn{T}he total magnetic-field amplification \jn{in a portion of the precursor} thus results from compression, and our simulation is uniquely suited to recover it.}}

\subsubsection{\jnr{Nonlinear plasma response}\label{plasma_turb}}

As the magnetic field grows to amplitudes $\langle\delta B/B_{\parallel 0}\rangle\ga 1$, plasma-density perturbations appear (Fig.~\ref{bell1}a for $x\ga x_3$), that are accompanied by velocity fluctuations. In this strongly nonlinear phase, plasma parcels begin to move, collide with each other \okr{and} collapse into turbulently moving filaments, causing distortions of the initially parallel magnetic modes.
The plasma also slows down in bulk, \mpo{in particular during the late phase. Whereas at time $t\gamma_{\rm max}= 13$ the electron-ion beam decelerated only from the initial drift speed of $0.4c$ to $\sim 0.35c$ in the region of strong turbulence ($x/\lambda_{\rm max}\ga 60$), by $t\gamma_{\rm max}= 26.9$ the \jn{beam} drifts slowly with $v\sim 0.15c$ ({\it red} lines in Fig.~\ref{profiles}d). Bulk CR motion with mean velocity of $\sim 0.1c$ is likewise observed, and the bulk velocities of the three plasma species converge \mpo{(the slight reduction in CR bulk speed toward $x\approx 180\,\lambda_{\rm max}$ results from the reflecting boundary condition for CRs)}. Further growth of the turbulence is thus limited in this region and the instability saturates.}

\mpo{If we envision Bell's instability to be an agent of magnetic-field amplification in the precursor region of SNR shocks, we must also consider it \jn{as} a means of shock modification by deceleration and compression of the upstream plasma \citep[for a review see, e.g.,][]{1987PhR...154....1B}. We initiate our simulations with a constant CR density, and so the backreaction is not driven by a gradient in CR pressure, but generates it. In any case, the feedback on the upstream plasma will also modify the thermal sub-shock, and a new balance between particle acceleration and its feedback will be established.
We do not explicitly model the shock that provides the CRs, and so we cannot determine that statistical equilibrium between shock acceleration of CRs, the build-up of magnetic turbulence in the precursor, and the backreaction of CRs and turbulence on the inflowing plasma.}

\subsubsection{\jnr{Electric field turbulence}\label{efield}}

Associated with the magnetic turbulence in the nonlinear stage is a turbulent electric field whose amplitude is an order of magnitude smaller than that of the magnetic field (see Fig.~\ref{profiles}b). \mpo{This electric field \mpof{initially} arises as $\bmath{v}\times\bmath{B}$ motional term of the bulk flow with
the mainly transverse magnetic fluctuations, and hence its dominant components are $E_y$ and $E_z$. The electric fluctuations appear when the magnetic modes have grown to nonlinear amplitude} (e.g., at $x/\lambda_{\rm max}\approx 45$ at $t\gamma_{\rm max}= 13$ in Fig.~\ref{profiles}b). 

Once the plasma attains turbulent motions, an $E_x$ component \mpof{appears} as well (for $x/\lambda_{\rm max}\approx 55$ at $t\gamma_{\rm max}= 13$, Fig.~\ref{profiles}b). In contrast to the perpendicular electric-field components that oscillate around zero, the average longitudinal component in the turbulent zone is essentially always positive, as demonstrated in Fig.~\ref{profiles}c. \mpo{Most of the $E_x$ field results from $v_yB_z$ and $v_zB_y$ terms.} 
\jn{The presence of this mean parallel electric field is a response of the system to the CR electric current \citep[e.g.,][]{bell05,zirakashvili08}}.
\jn{Its role} in the microscopic particle motions will be discussed in Section~\ref{micro}.     

\begin{figure}
\centering
\includegraphics[width=\linewidth]{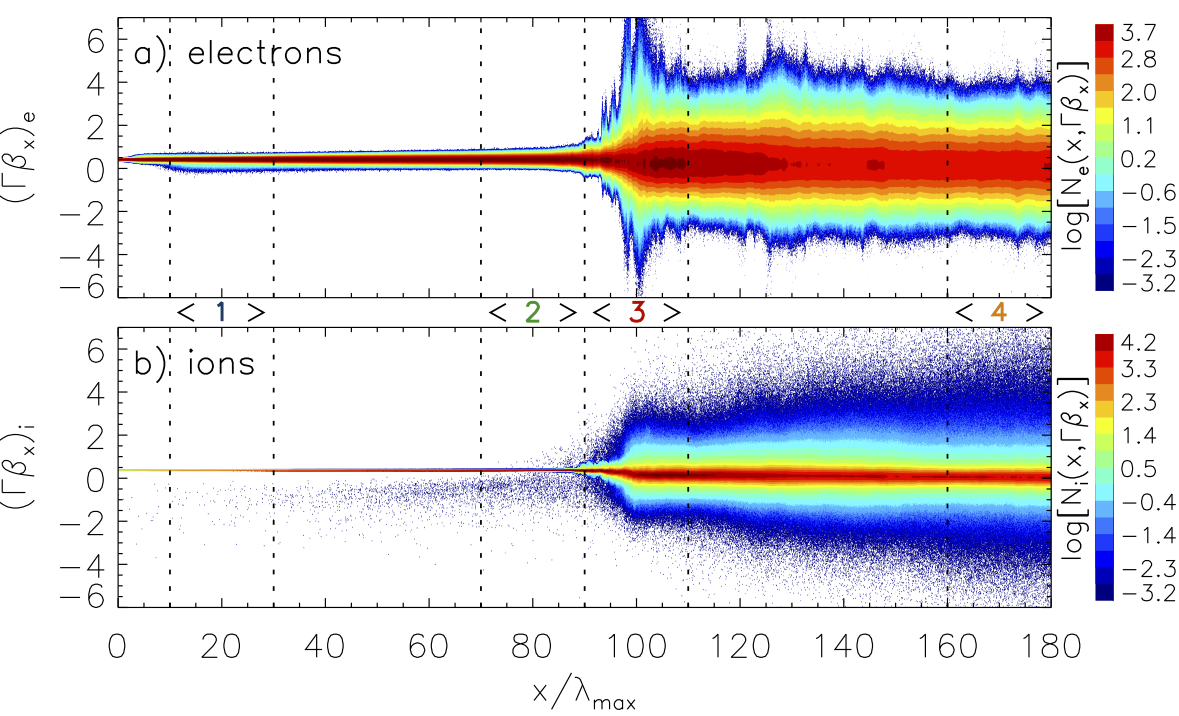}
\caption{Longitudinal phase-space distribution of plasma electrons (a) and ions (b) at the end of the simulation run at  $t\gamma_{\rm max}=26.9$. The four regions marked with vertical dotted lines and colour-coded numbers denote portions of the box from which the particle spectra in Figs.~\ref{specr}, \ref{specele} \okr{and} \ref{specion} are extracted.} 
\label{phase}
\end{figure}

The spatial structure of the plasma velocity profile can be traced in Fig.~\ref{phase} that shows the longitudinal phase-space distribution at $t\gamma_{\rm max}= 26.9$ (see also discussion on Fig.~\ref{crtraj} in Sec.~\ref{micro}). \mpo{Evident in Fig.~\ref{phase} for $x/\lambda_{\rm max}\ga 90$ is the significant plasma heating \jn{in the region of} the strong density and velocity fluctuations. The irregular phase-space structure for electrons visible in Fig.~\ref{phase}a reflects localized 
heating, the details of which will be discussed in the next section.} Here we only note that although the dominant part of the electric fields arises from magnetic-field transport, there also exists an electrostatic field component. It can be calculated by either performing a Lorentz transformation of the fields into the local plasma rest frame or by directly solving the Poisson equation. Both methods reveal electric-field fluctuations that can inelastically scatter particles and modify their distribution.   

\subsubsection{\jnr{Phase-space structure of the compression wave}\label{phase-space}}


\okr{As noted above, our realistic precursor modeling reveals the compression wave that can be observed in Figs.~\ref{bell2} and \ref{profiles} for $x/\lambda_{\rm max}\approx 90-130$. \jnr{The phase-space structure in the vicinity of this feature is shown} in Fig.~\ref{phase}.} \jnr{Beside the density increase by a factor of four (Fig.~\ref{profiles}a), the plasma is strongly heated,} as observed in nonrelativistic shock waves.
The structure \jnr{also} exhibits \jnr{a} distinctive feature of a supercritical shock -- in the upstream region 
a population of the reflected plasma ions is observed that it is clearly visible in Fig.~\ref{phase}b as a diffuse particle beam with negative momenta for $x/\lambda_{\rm max}\la 90$, extending almost to the plasma injection boundary at the left side of the simulation box. Although the phase-space density of the reflected ions is very low, they \mpo{cause substantial heating of the incoming plasma} electrons. They possibly also modify the magnetic-field structure through, e.g., ion-beam streaming instabilities \citep[e.g.,][]{niem12}. Any distortions of the magnetic turbulence resulting from such instabilities would appear as magnetic filaments and are \jn{thus} difficult to identify in the fluctuations that are already organized in the form of filaments upstream of the \jnr{shock-like structure}. Additional effects of the \jnr{compression wave} will be discussed in particle spectra presented in the section that follows.  

\subsection{Microphysics of the saturation process\label{micro}}

The energy and momentum transfer between CRs and the plasma that leads to the saturation of the nonresonant instability is mediated by highly turbulent large-amplitude electromagnetic fields. These fields scatter the particles and significantly modify their distributions. 
\jnr{In this section we describe these processes at the micro-physical level. CR scattering in the Bell's turbulence is presented in Section~\ref{crscatt} and discussed in the context of their spatial diffusion in Section~\ref{crdiffusion}. The background electron-ion plasma heating and scattering is described in Sections~\ref{amspectra}, \ref{amicro} and~\ref{ionscatt}.}

\subsubsection{Cosmic-ray scattering\label{crscatt}}

\begin{figure}
\centering
\includegraphics[width=\linewidth]{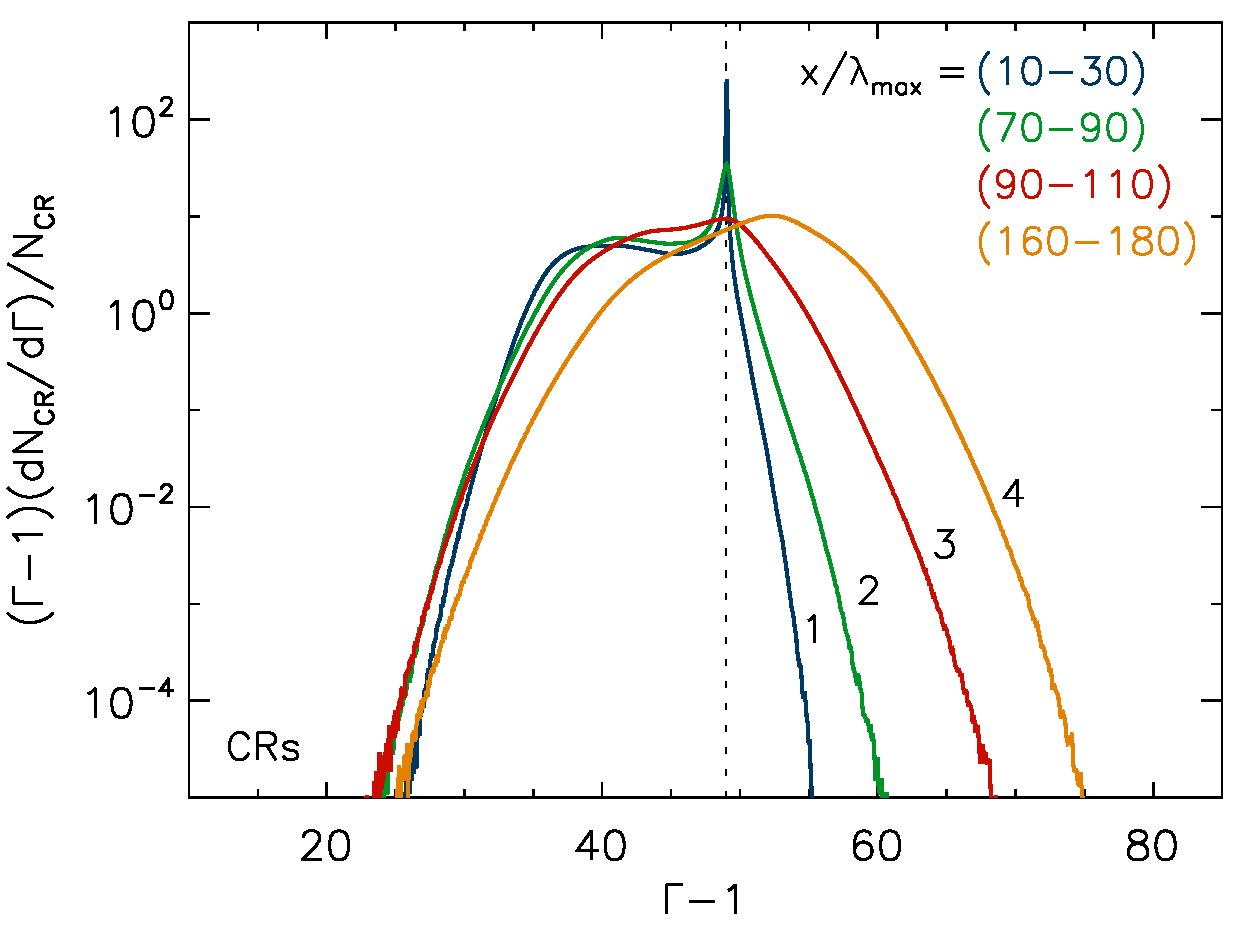}
\caption{Kinetic-energy spectra of CR ions in the four colour-coded regions shown in Fig.~\ref{phase}. 
The spectra are normalized and calculated at time $t\gamma_{\rm max}=26.9$ in the simulation frame. The vertical dotted line at $\Gamma-1=49$ marks the initial mono-energetic CR distribution.} 
\label{specr}
\end{figure}

\begin{figure}
\centering
\includegraphics[width=\linewidth]{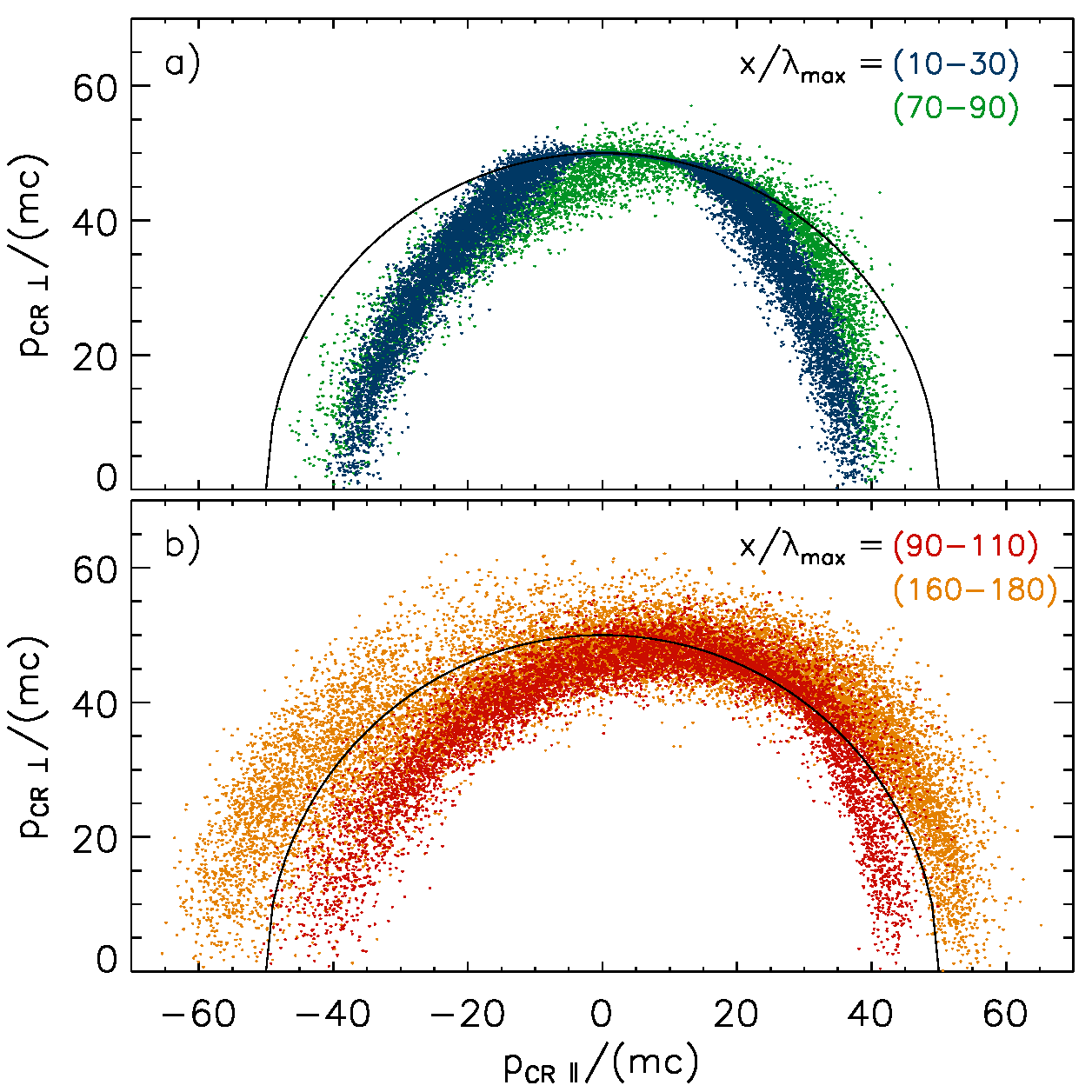}
\caption{Phase-space distributions of CRs in the simulation frame and at time $t\gamma_{\rm max}=26.9$, 
compared with the initial distribution marked with a thin semicircle. 
\mpo{As in Fig.~\ref{specr}, we distinguish by colour the distributions in four} regions marked in Fig.~\ref{phase}.} 
\label{craniso}
\end{figure}

Figs.~\ref{specr} and \ref{craniso} show CR kinetic-energy spectra and phase-space distributions at the end of our simulation at $t\gamma_{\rm max}=26.9$. 
Results for \jnr{the four} physically distinct regions marked in Fig.~\ref{phase} are presented with different colours. The degree of CR scattering and anisotropy is apparent. It is also evident that most of the scattering is inelastic and results in either acceleration or deceleration of CR particles as they probe the turbulent precursor regions.  

Deeper insight into CR scattering is provided by
Fig.~\ref{crtraj}, that shows the kinetic-energy evolution and the $x$-component of particle trajectories for typical CRs. 
Scattering in energy commences at time $t\gamma_{\rm max}\approx t_3=8$ (see Fig.~\ref{avmag}), 
when the amplitude of magnetic turbulence becomes comparable to the strength of the homogeneous field, 
$\langle\delta B/B_{\parallel 0}\rangle\ga 1$. 
The mean energy increases from the initial $\Gamma-1=49$ to approximately $50$ by $t\gamma_{\rm max}\approx 18$ and shows little change at later times. Much of the mean energy gain arises from the bulk energy transfer from the ambient plasma beam that is demonstrated in Fig.~\ref{profiles}d.
Correspondingly, the most efficient scattering takes place \mpof{before} $t\gamma_{\rm max}\approx 18$. Inspection of individual particle trajectories shows that CRs initially move on almost rectilinear paths on account of their large Larmor radii and then become deflected in pitch angle as they pass through regions of strong magnetic field of \mpof{varying} polarity.    
\mpo{Part of the scattering may be resonant with the waves generated via the instability \citep{2009ApJ...706...38S,luo}, but most of it results from inelastic scattering off 
\jn{electromagnetic turbulence.}
Consequently the time histories of particle energy in Fig.~\ref{crtraj}a show corrugated structures. 
Scattering in energy occurs in the electric fields associated with the motion} of turbulent magnetic field. 

\begin{figure}
\centering
\includegraphics[width=\linewidth]{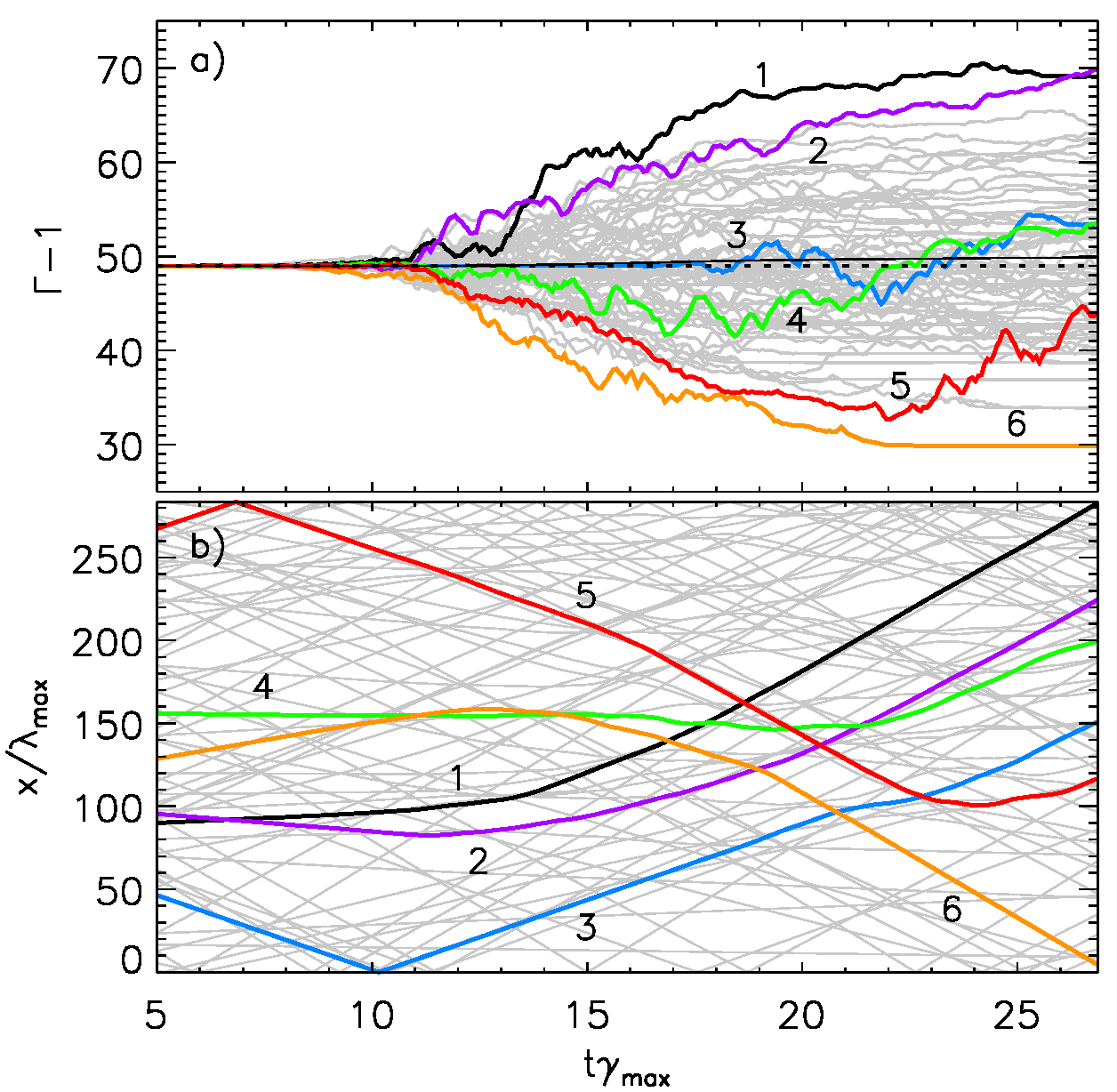}
\caption{Time evolution of the kinetic energies (a) and $x$ coordinates (b) for a sample of cosmic-ray particles. Selected particles are \okr{marked with numbers from 1 to 6 and} highlighted with colour for better visibility. Kinks in the trajectories are due to particle reflections off the box boundaries. The dotted horizontal line at $\Gamma-1=49$ marks the initial CR energy. The \okr{thin} solid black line shows the evolution of the mean energy for all CR particles in the simulation.} 
\label{crtraj}
\end{figure}

\mpo{An interesting feature can be observed in Fig.~\ref{crtraj}: CRs that travel along with the plasma flow in the positive $x$-direction are accelerated (e.g., particles \okr{No. 1 and 2,} shown with {\it black} and {\it magenta} lines), whereas those moving against the flow become decelerated (e.g., CR \okr{No. 6,} shown with the {\it yellow} line). Particles that change the $x$-component of their motions upon scattering seem to gain or \okr{lose} their energy accordingly \jn{(e.g., CRs \okr{No. 3, 4, and 5,} shown with \textit{blue}, \textit{green} \okr{and} \textit{red} lines)}. To be scattered, a particle must reside in the region of strongly nonlinear turbulence, and hence fluctuations in energy are seen predominantly at times $t\gamma_\mathrm{max}\gtrsim 10$ \jn{and particles at} locations $x \gtrsim 80 \lambda_\mathrm{max}$. These characteristics of CR scattering result from the structure of turbulent electric fields described in Section~\ref{efield}.}  

The $E_x$ field component in the turbulence region is on average nonzero and directed in the positive $x$-direction. 
\jn{It is induced in the system by the CR electric current and allows bulk-energy transfer between the CRs, the ambient plasma \okr{and the} turbulence. In the initial CR rest frame of our simulation the mean electric field will extract energy from the ambient plasma beam at a rate $-E_x j_{\rm ret}$, causing a bulk acceleration of CRs and a bulk slow-down of the ambient beam flow velocity, as observed in Figs.~\ref{profiles}d and \ref{crtraj}a. At the level of individual particle trajectories the same field causes}
a net acceleration for CR ions traveling along the flow and a net deceleration for CR particles moving in the negative $x$-direction. \mpo{Momentum change by the dominant $E_y$ and $E_z$ fields averages to zero for CRs}. 
The \mpo{thermal speed of ions is small, as is their Larmor radius, and the ions are more susceptible to the effects of the turbulent fields than are CR particles. We will  describe in detail the microphysics of ion scattering and their heating in \jnr{Sections~\ref{amspectra} -- \ref{ionscatt}.}

\mpo{The yellow line in Fig.~\ref{specr} indicates a high-energy bump resulting from efficient CR scattering, which requires a large region of strong turbulence. Evidently our simulation setup provides a sufficient volume filled with turbulence, but in reality, this region may be limited in size by the presence of a shock and also by lower-energy CRs that reside} closer to the shock and modify the turbulence structure there. Hence, the CR spectra formed in this region are presented here to illustrate the main features of the physical processes in the precursor \okr{and} not to provide quantitative predictions. The latter would require a full shock simulation, including the electron dynamics. A similar remark applies to the spectrum and phase-space distribution of CRs calculated for the very far upstream region 
($x/\lambda_{\rm max}=10-30$, {\it blue} line/points in Fig.~\ref{specr}/Fig.~\ref{craniso}\jnr{a}). In this zone \mpo{CRs scattered off their initial distribution might be overabundant on account of the reflecting boundary that prohibits leakage of these particles away from the strongly turbulent region in which they originated. A similar caveat applies to the region $x/\lambda_{\rm max}=160-180$ where reflected CRs from the rear boundary lower the bulk speed in the simulation.
The distributions calculated in the zone of small-amplitude turbulence ($x/\lambda_{\rm max}=70-90$, {\it green} line/points) and at the compression wave 
($x/\lambda_{\rm max}=90-110$, {\it red} line/points)
are least influenced by boundary conditions. Some anisotropy in these distributions comes from particles reflected off the compressed field of the shock-like structure.}

\subsubsection{\jnr{Spatial diffusion of cosmic rays}\label{crdiffusion}}

\begin{figure}
\centering
\includegraphics[width=\linewidth]{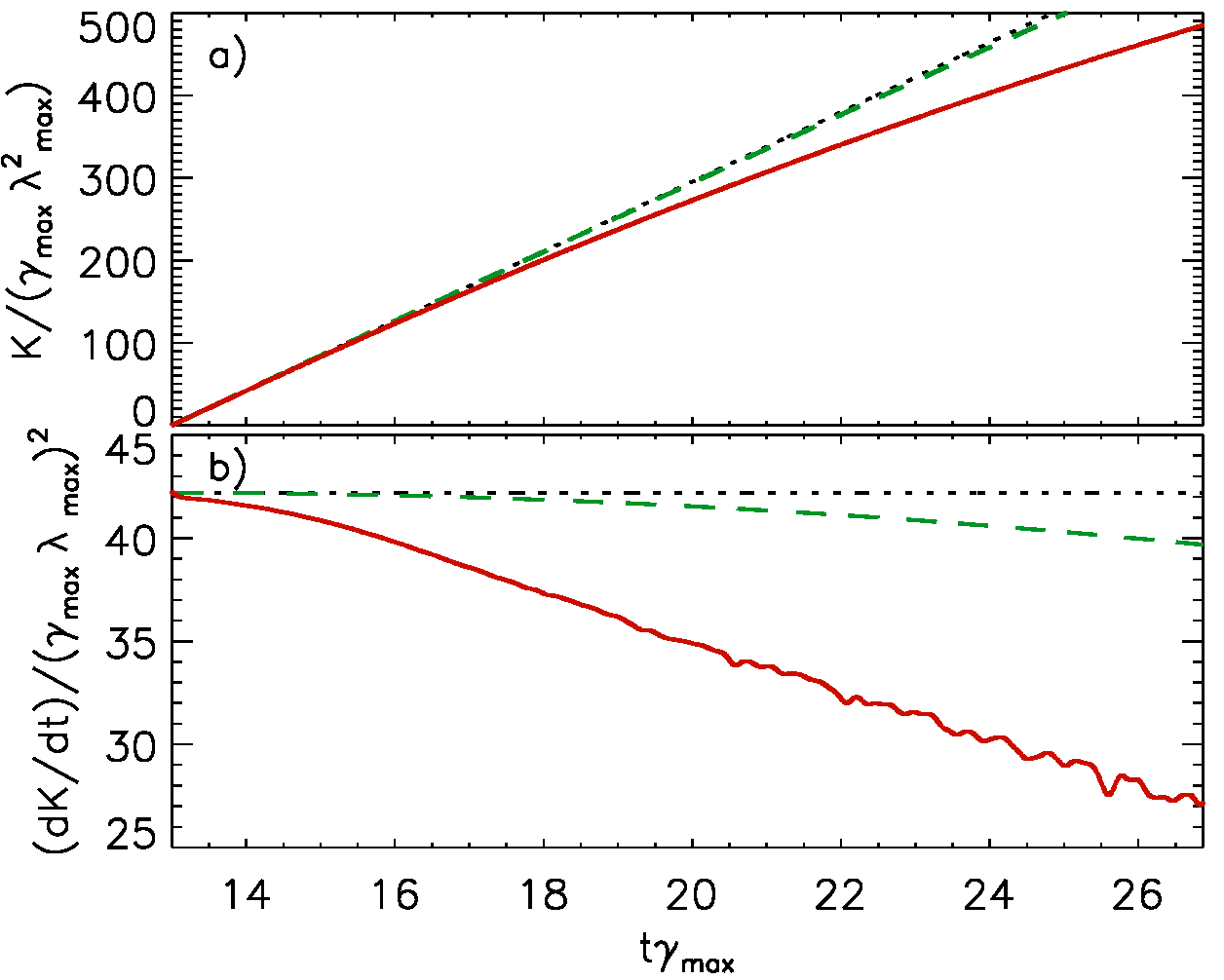} 
\caption{\okr{Solid \emph{red}} lines display the spatial diffusion coefficient, $K(t)$, (a) and its time derivative (b) as function of time, calculated for CRs residing in the region of strong magnetic turbulence, $x/\lambda_{\rm max}\ga 110$, from time 
$t\gamma_{\rm max}=t_4=13$ (see Fig.~\ref{avmag}) until the end of the numerical experiment. \okr{Dashed \emph{green}} lines correspond to the motion of particles in the initial homogeneous magnetic field. Extension of their initial directions is shown with dotted lines.} 
\label{crdiff}
\end{figure}

The motion of CRs in the region of strong magnetic turbulence resembles diffusion in space. 
\mpo{We can calculate a running diffusion coefficient as}
\begin{equation}
K(t)=\frac{\langle(x-x_0)^2\rangle + \langle(y-y_0)^2\rangle}{4\,t},
\end{equation}
where $x_0$ and $y_0$ are the initial particle locations. The \mpof{running} diffusion coefficient, $K(t)$, and its time derivative are shown in Fig.~\ref{crdiff} with \okr{solid} \textit{red} lines. The coefficient is calculated from time 
$t\gamma_{\rm max}=t_4=13$ until the end of the \mpo{simulation and \mpof{for} all CRs that continuously reside in the strongly turbulent zone at $x/\lambda_{\rm max}\ga 110$. Particles diffusing in from the small-amplitude region, as well as those that leak into the latter zone, \jn{are} ignored.} 

One notes that CR motion is not truly diffusive, as the \mpo{running diffusion coefficient does not reach constancy, \(K(t)={\rm const}\). \okr{However, i}t clearly deviates from helical motion. 
If particles would only gyrate in the initial homogeneous magnetic field, then the running diffusion coefficient would behave as}
\begin{equation}
\frac{K(t)}{\lambda^2_{\rm max}\gamma_{\rm max}}=20.8\,\gamma_{\rm max}t+
\frac{10^5}{3\,\gamma_{\rm max}t}\,\sin^2\left(\frac{\gamma_{\rm max}t}{40}\right)\ ,
\label{Diff}
\end{equation}
which leads to the \okr{dashed} \textit{green}
\mpo{lines in Fig.~\ref{crdiff}. $K(t)$ clearly deviates from that but does not reach a constant level during the simulation. As Bell's instability operates at $k\,r_\mathrm{CRg}\gg 1$, the magnetic-field fluctuations can only impose scattering, and the time derivative of $K(t)$ in Fig.~\ref{crdiff}b indicates a linear decrease toward $dK(t)/dt=0$, whereas gyration would lead to a parabolic decline. One can estimate the time, $t_{\rm diff}$, needed for CRs to transition into the diffusion regime by extrapolating the derivative down to $dK/dt=0$, i.e., $K(t_{\rm diff})={\rm const}$. The required time is $t_{\rm diff}\gamma_{\rm max}\approx 36.5\pm 1$, and the spatial diffusion coefficient would reach}
$D=K(t_{\rm diff})=(770\pm 20)~\lambda^2_{\rm max}\gamma_{\rm max}$.
The time $t_{\rm diff}$ considerably extends beyond the duration of our numerical experiment.

\mpo{Let us compare our approximate value of the diffusion coefficient, $D$, with that allowed in the Bohm limit. For that purpose we shall write the Bohm diffusion \mpof{coefficient} as 
\begin{equation}
D_{\rm B}=\frac{r^2_{\rm CRg,f}}{\tau},
\end{equation}
where $r_{\rm CRg,f}$ is the final CR gyro-radius in \jn{the amplified magnetic field} \okr{and} $\tau$ is the mean time between scattering events. In the \jn{root-mean square field}, the final CR gyro-radius is 
$r_{\rm CRg,f}\approx 18\lambda_{\rm max}$. 
The typical coherence length of the magnetic turbulence is about 
$7\lambda_{\rm max}$, and the time needed for CRs to propagate one coherence length may be taken as proxy of the scattering time,}
$\tau\gamma_{\rm max}\approx 0.44$. 
Our estimate of the Bohm diffusion coefficient hence is
\begin{equation}
D_{\rm B}\approx 740~\lambda^2_{\rm max}\gamma_{\rm max},
\end{equation}
which is numerically similar to our estimated value of $D$. One can thus conclude that the spatial CR scattering observed in the strongly turbulent region generated due to the nonresonant instability is compatible with \mpo{Bohm diffusion \jn{in the amplified magnetic field}. Note, that similar conclusion was achieved for pitch-angle scattering of a mildly relativistic CR beam interacting with} Bell's turbulence \citep{niem10}.  

\subsubsection{\jnr{Spectra of ambient plasma} \label{amspectra}}


\begin{figure}
\centering
\includegraphics[width=\linewidth]{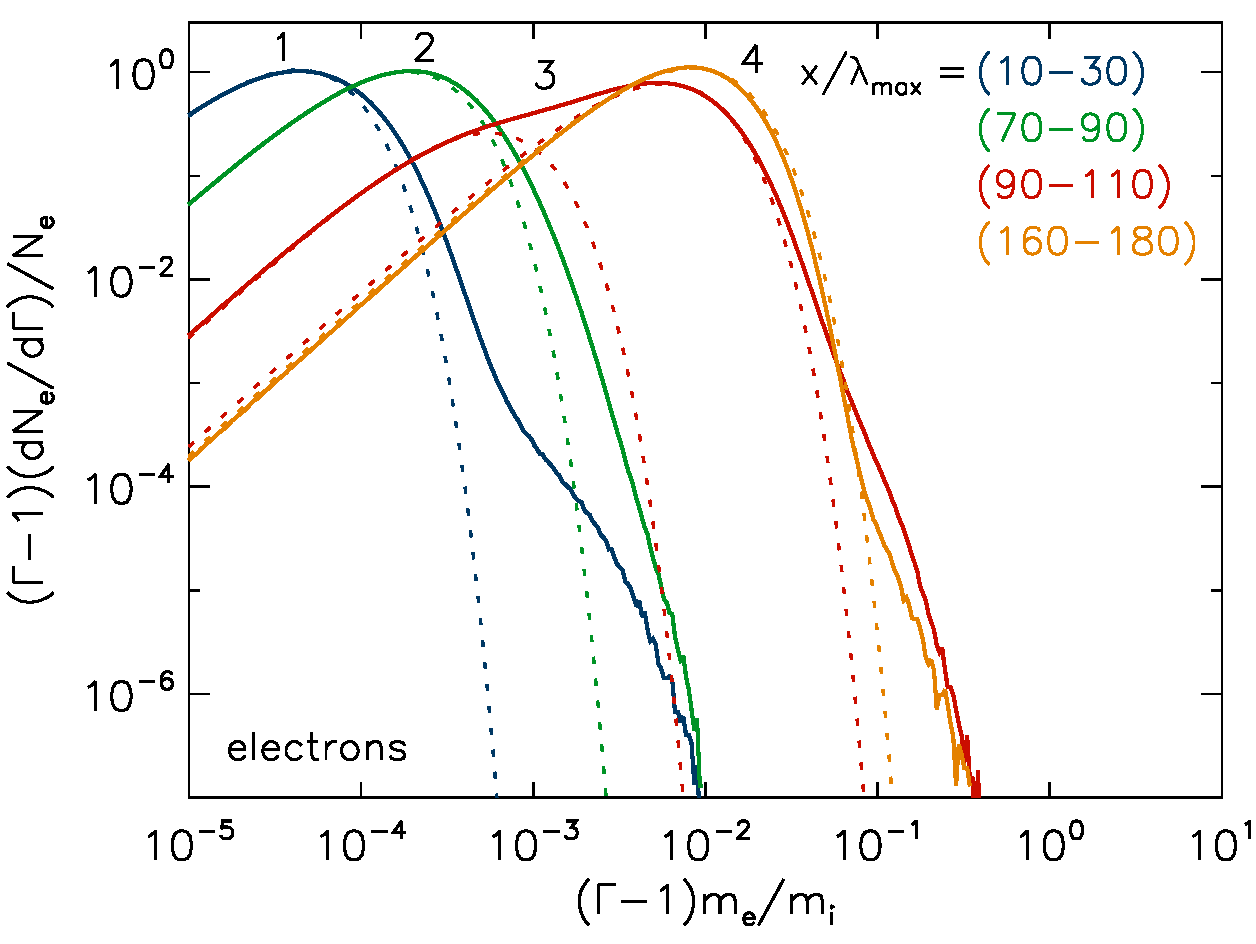}
\caption{\mpo{Kinetic-energy spectra of plasma electrons in the four regions along the flow marked in Fig.~\ref{phase}. The spectra are normalized and calculated at time $t\gamma_{\rm max}=26.9$ in the \emph{local} plasma frame. The electron energy is given in units of the ion rest-mass energy.} Dotted lines indicate fits of a relativistic Maxwellian.} 
\label{specele}
\end{figure}

\begin{figure}
\centering
\includegraphics[width=\linewidth]{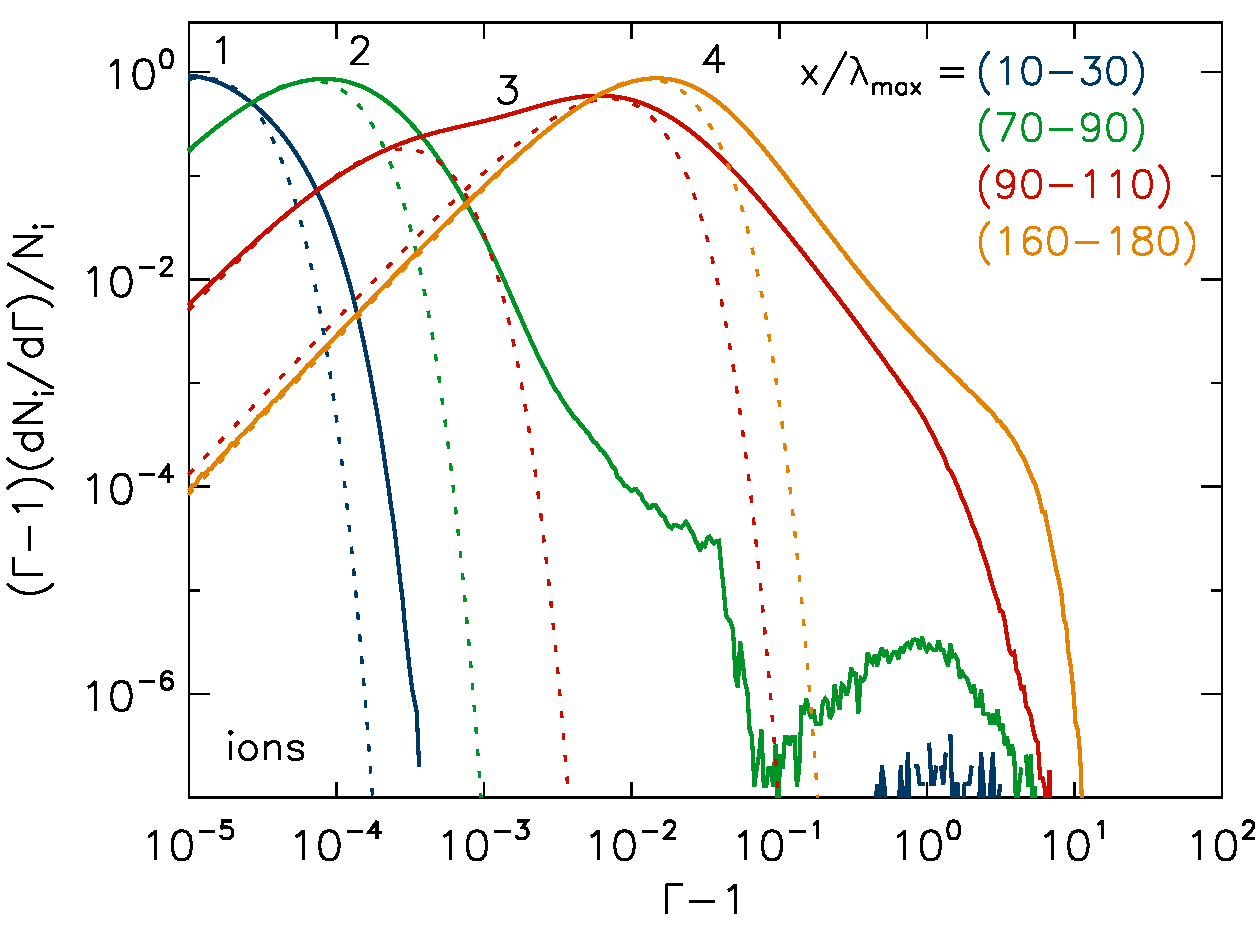}
\caption{Kinetic-energy spectra of plasma ions in four regions across the flow (see Fig.~\ref{specele}).} 
\label{specion}
\end{figure}

\mpo{We extract particle data for the four regions along the flow, that are marked in Fig.~\ref{phase}, and compute kinetic-energy spectra for electrons and ions at the end of the simulation run at time $t\gamma_{\rm max}=26.9$, that are shown in Figs.~\ref{specele} and \ref{specion}, respectively. The spectra are calculated in the local plasma rest frame, i.e., accounting for turbulent plasma motions. 

The plasma particles are successively heated as they stream through the background of CRs, exciting the nonresonant instability that nonlinearly grows in the region of strong turbulence. Effective plasma heating is evident in the band $x/\lambda_{\rm max}=90-110$ (\textit{red} lines, \okr{No.~3}), and the electron spectra have largely relaxed to a Maxwellian in the area $x/\lambda_{\rm max}=160-180$ (\textit{yellow} lines, \okr{No.~4}).} The ion spectra still clearly deviate from Maxwellians and show supra-thermal tails. The two species remain close to equipartition in bulk, but a substantial fraction of ions can achieve more than an order of magnitude higher energies than do electrons. This suggests that \jnr{al}though the heating mechanism for both plasma species \mpof{appears} similar, \mpo{either thermal relaxation of ions is slow or} there exists another process that enables further energization of ions beyond the thermal pool. 

\begin{figure*}
\centering
\includegraphics[width=0.49\linewidth]{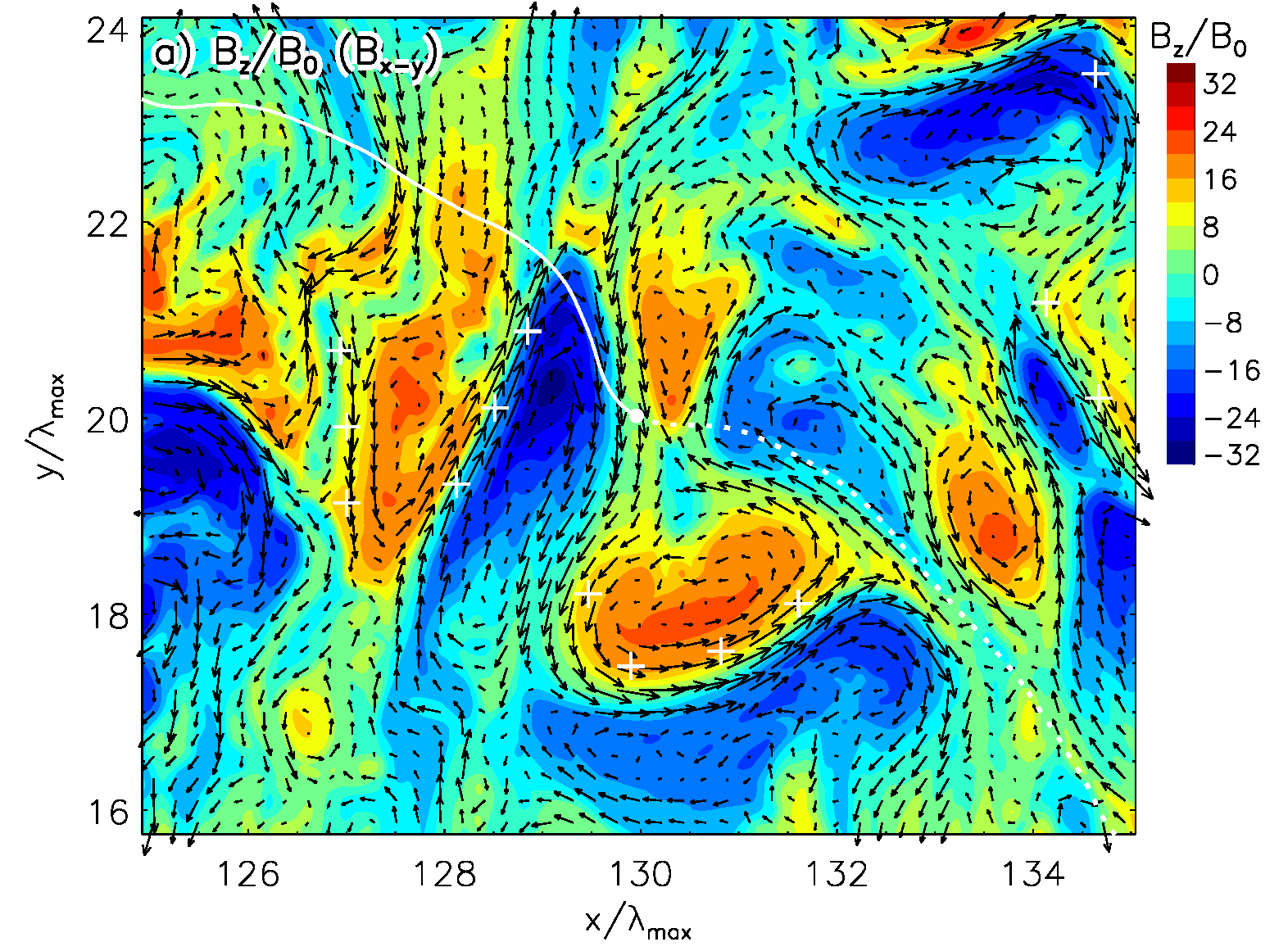}
\includegraphics[width=0.49\linewidth]{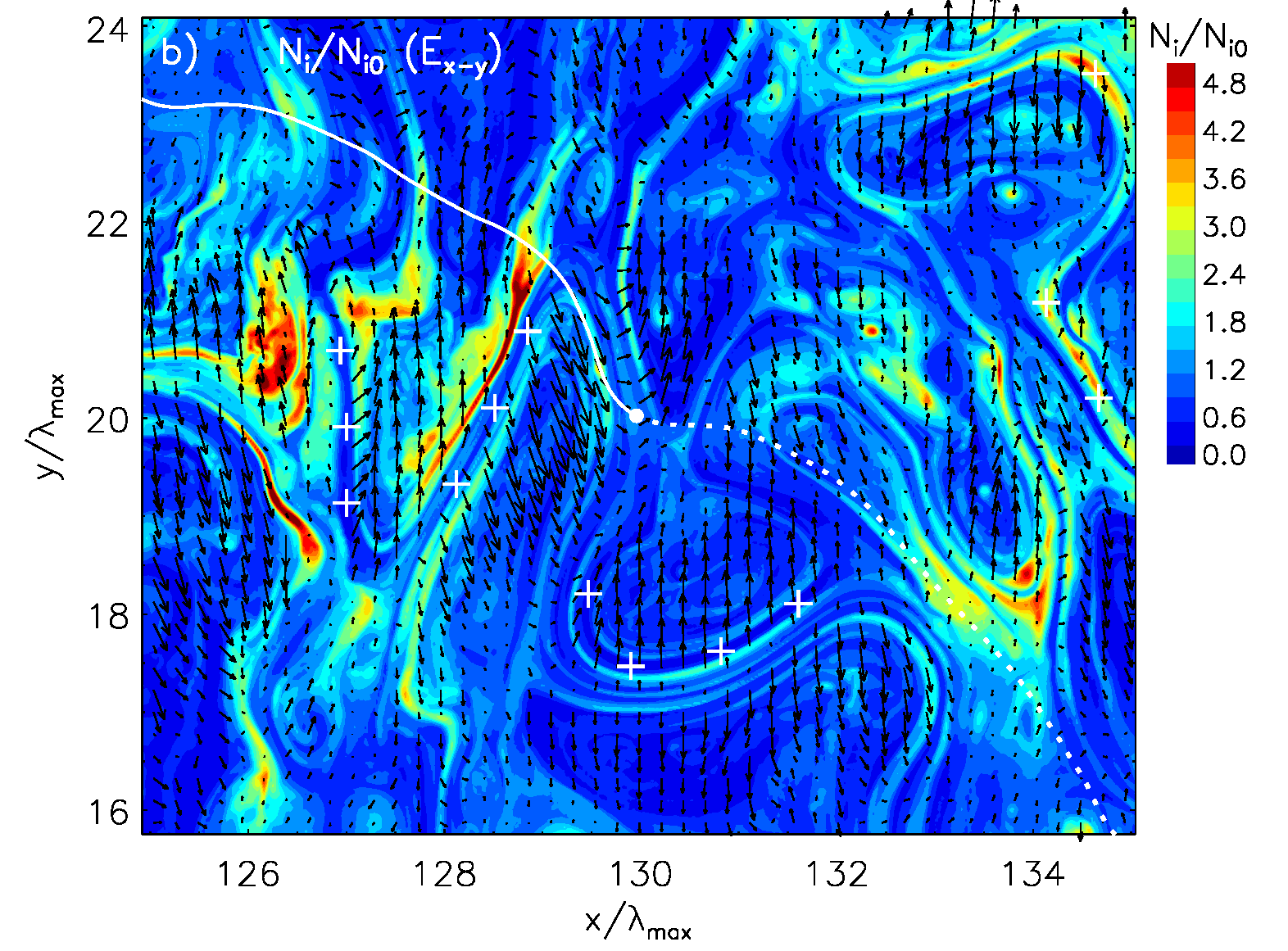}
\includegraphics[width=0.49\linewidth]{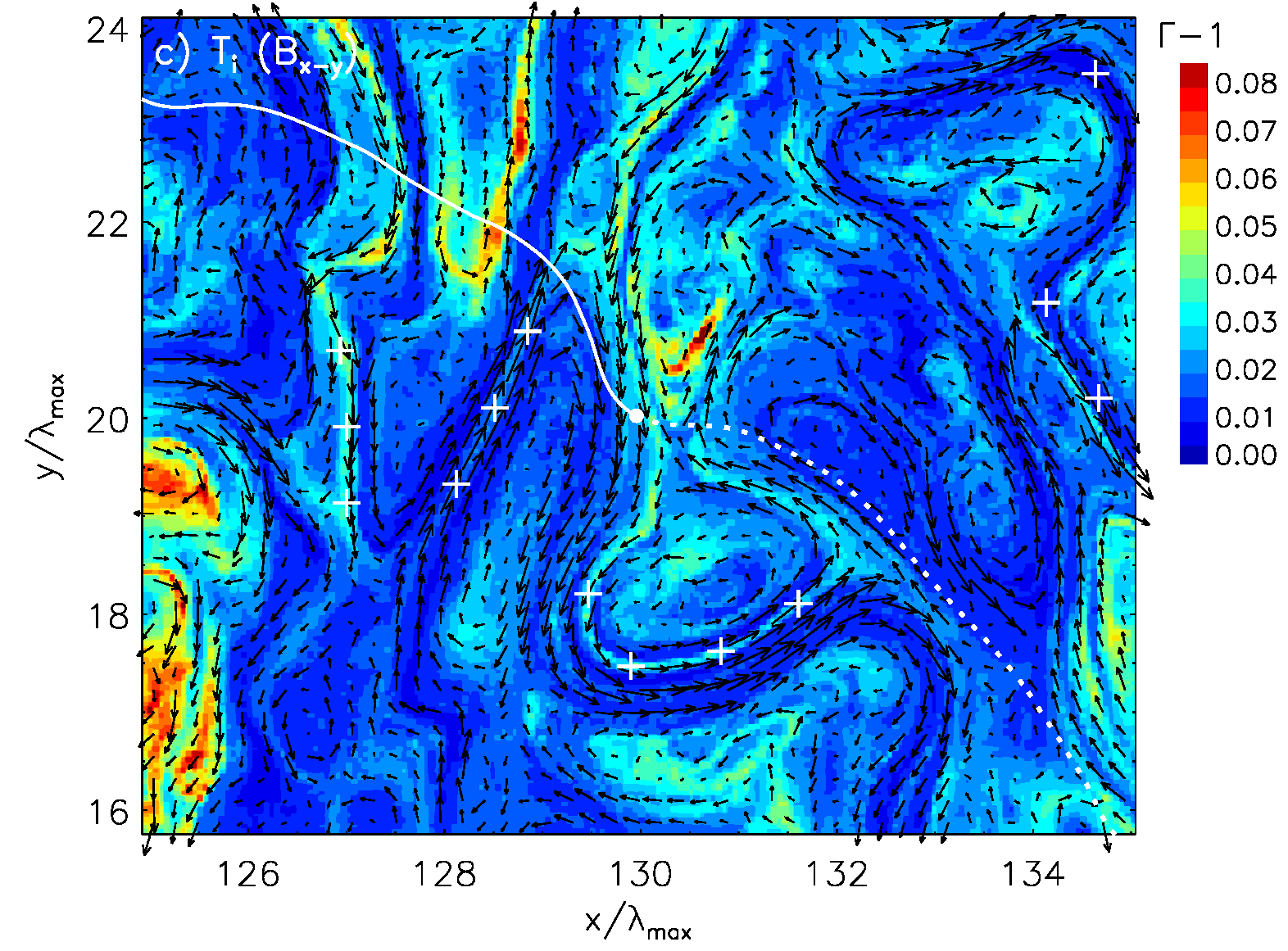}
\includegraphics[width=0.49\linewidth]{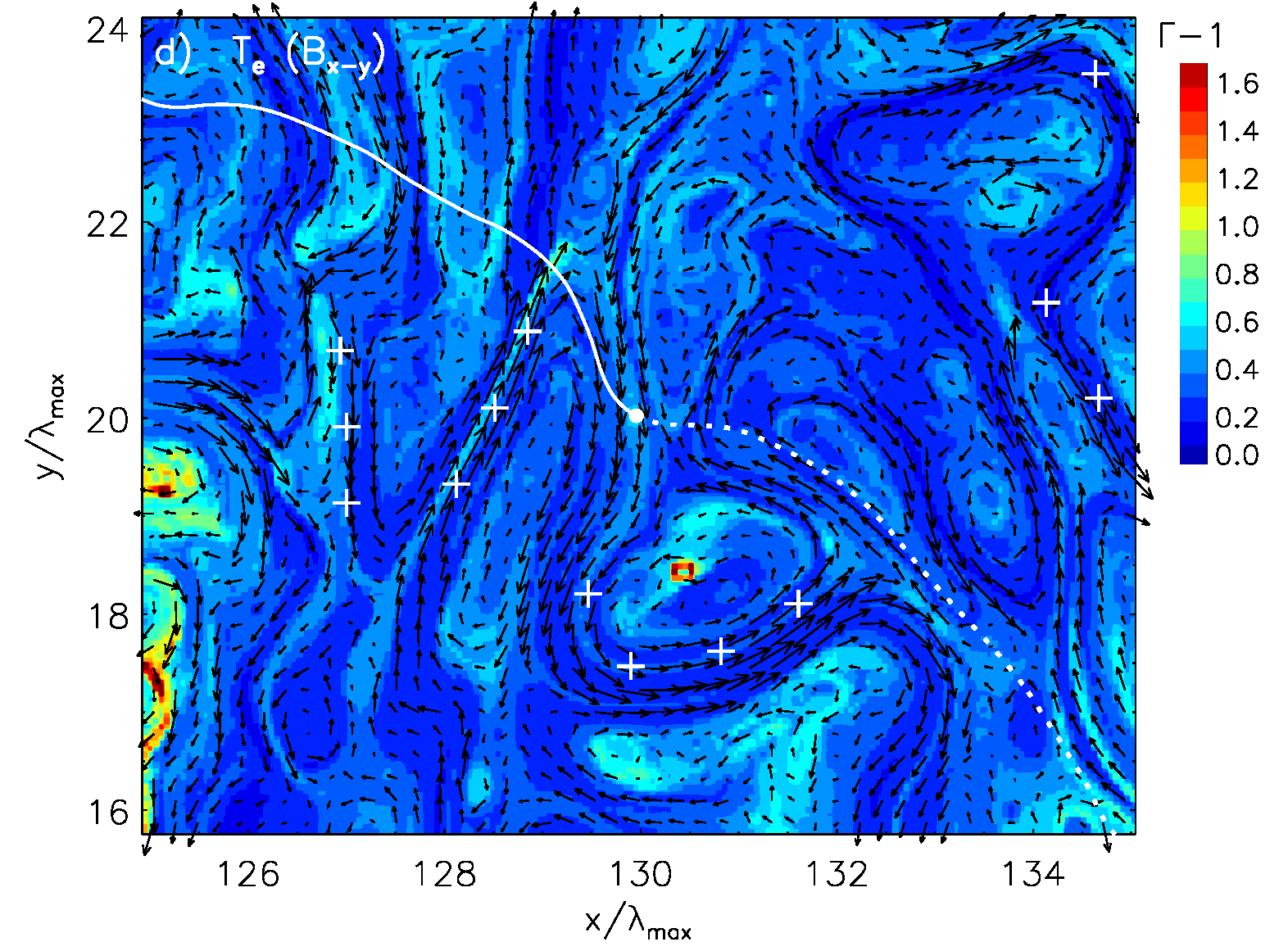}
\caption{\mpo{Distribution of the normalized magnetic-field amplitude $B_z/B_0$ (a), normalized ion \jnr{number} density $N_i/N_{i0}$ (b) \okr{and} the mean kinetic energy (temperature) of ions (c) and electrons (d), all taken at time $t\gamma_{\rm max}=15$ 
in a small region harboring strong turbulence. Linear scales are used in all maps. Overlaid on the maps in panels (a), (c) \okr{and} (d) are in-plane magnetic-field lines \okr{and} likewise in-plane electric-field lines in panel (b). The white crosses are intended to facilitate cross-correlation of various details. One example of an ion trajectory is shown as white solid line for times $t\gamma_{\rm max}\le 15$ and as dashed line for later times.}} 
\label{turb}
\end{figure*}

\begin{figure*}
\centering
\includegraphics[width=0.49\linewidth]{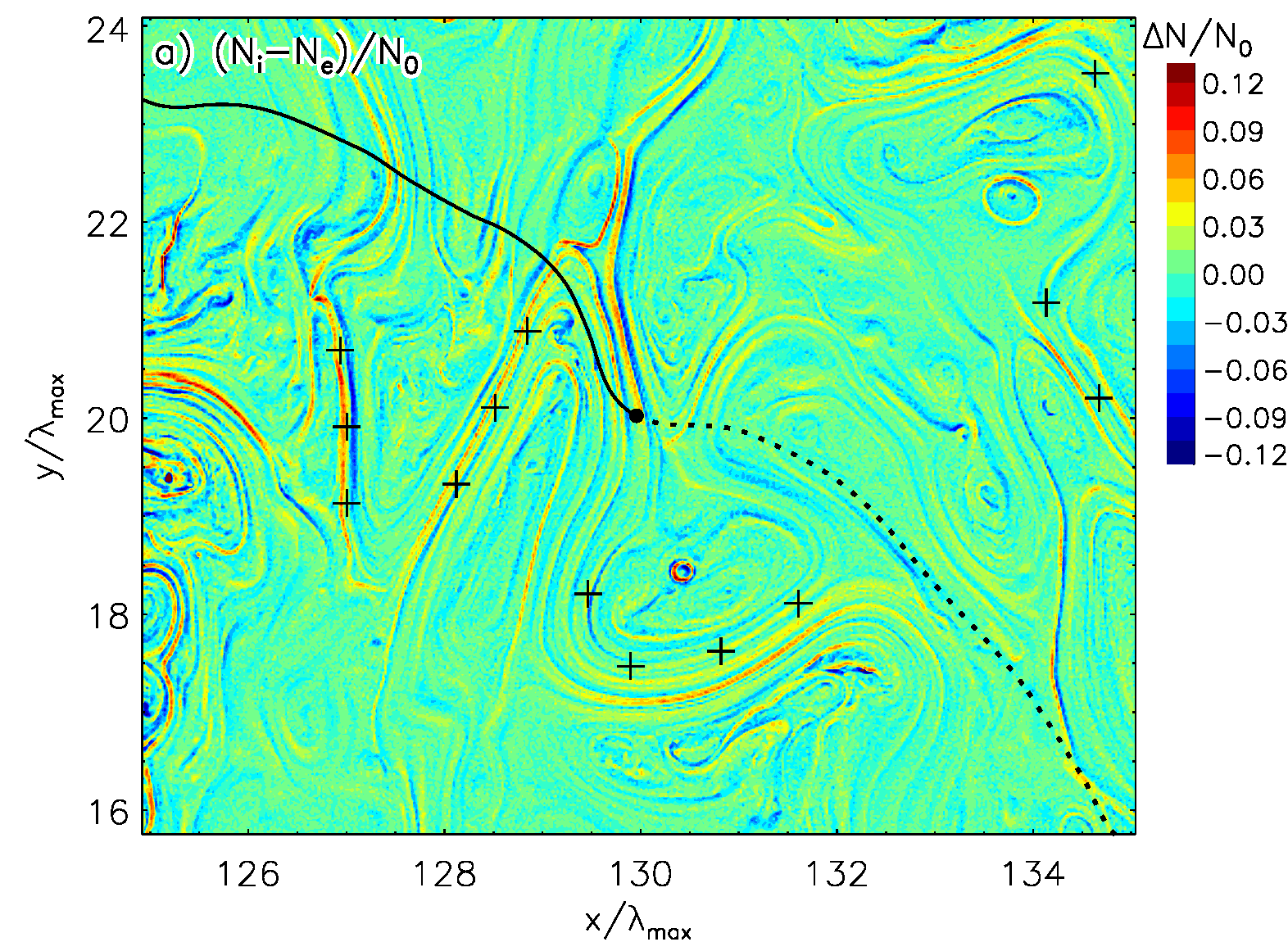}
\includegraphics[width=0.49\linewidth]{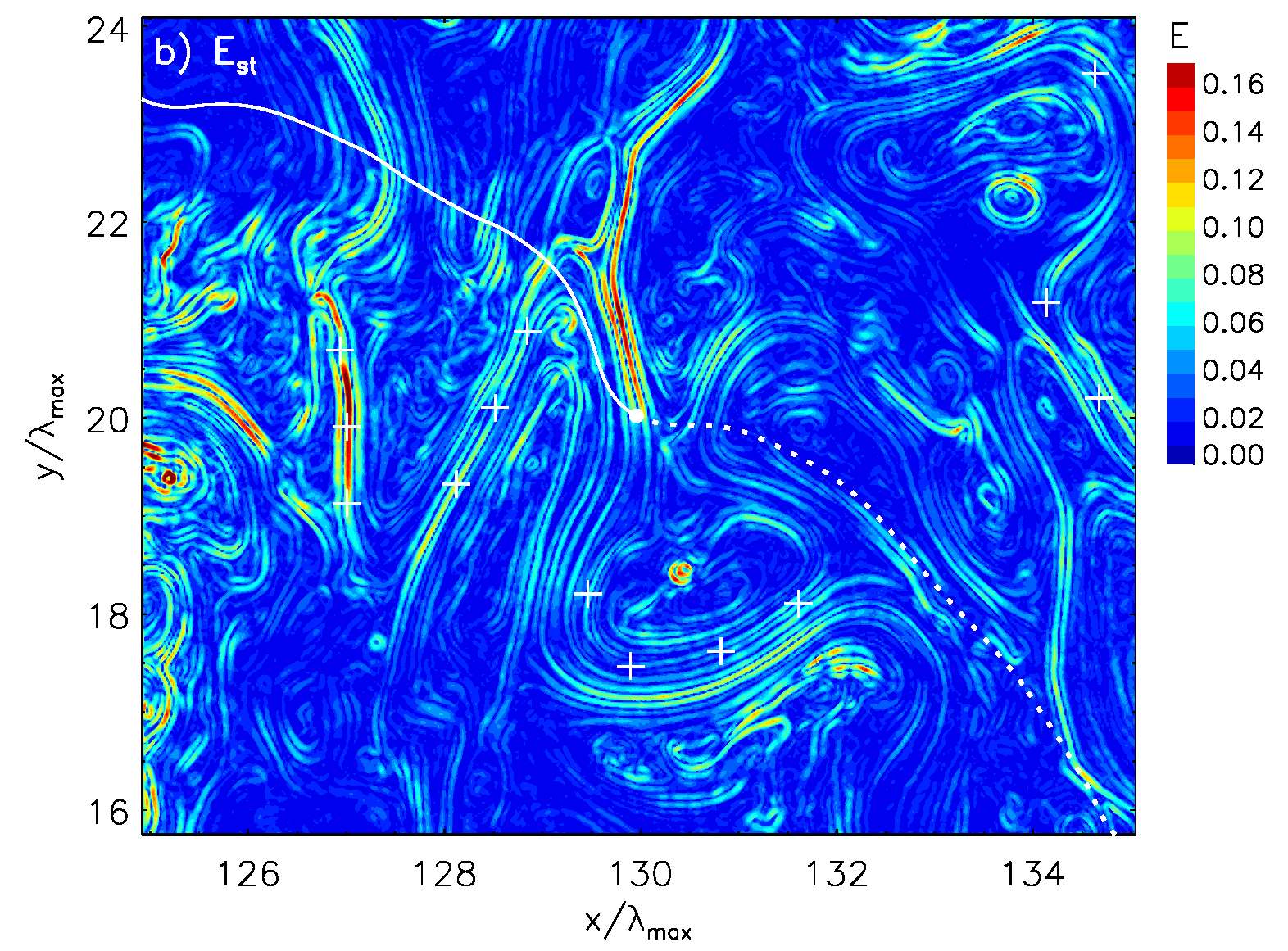}
\caption{\mpo{Distribution of the charge density in the plasma (a) and the resulting electrostatic} field (b).} 
\label{epot}
\end{figure*}

The spectra calculated in the region of the compression wave ($x/\lambda_{\rm max}=90-110$, \textit{red} lines, \okr{No.~3}), through which the system transitions from small-amplitude \mpo{fluctuations to highly nonlinear turbulence, show a complex structure, as expected. As discussed already in \jnr{Section~\ref{phase-space}}, plasma compression causes reflection of ions back to the upstream region of the compression zone (compare Fig.~\ref{phase}). The reflected population can be observed as a high-energy bump in the ion spectrum for the region $x/\lambda_{\rm max}=70-90$ (\textit{green} line \okr{No.~2} in Fig.~\ref{specion}). These particles interact with the incoming plasma and cause a tail in the otherwise thermal ion distribution that covers the energy range  $5\cdot 10^{-3}\ga\Gamma_i-1\la 10^{-1}$. 
The reflected particles reach far upstream and are \mpof{still} \jn{visible} in the spectra for the area $x/\lambda_{\rm max}=10-30$ (\textit{blue} line \okr{No.~1} in Fig.~\ref{specion}). The counterstreaming ions also influence the electrons that react with the formation of spectral tails} (\textit{blue} and \textit{green} lines in Fig.~\ref{specele}).

\subsubsection{\jnr{Micro-physics of plasma heating} \label{amicro}}
To investigate the micro-physics of plasma heating we analyze the turbulence structure in a small portion of the computational box located in the region of strongly nonlinear field fluctuations.  
Fig.~\ref{turb} shows maps of the $B_z$ magnetic-field amplitude and the ion \jnr{number} density, overlaid on the in-plane magnetic and electric field lines, respectively (Fig.~\ref{turb}a-b). We also display maps of the mean kinetic energy of ions and electrons (Fig.~\ref{turb}c-d) that serve as proxy for their temperatures. The maps are calculated at time $t\gamma_{\rm max}=15$ and are representative of the strongly nonlinear stage of \mpo{turbulence development. To aid the cross-correlation of features in the maps we add crosses as marks.}

\mpo{To be noted from the figure is the organization of the turbulent plasma into filaments and islands of enhanced density that surround plasma voids of very low density (Fig.~\ref{turb}b). Inspection of the plasma velocity fluctuations reveals that these structures are vortices that are filled \mpof{with} very strong perpendicular magnetic field ($B_z$) and wrapped in quasi-circular in-plane magnetic field ($B_x$ and $B_y$).}
These features are known from earlier studies of the nonresonant instability and have been explained as consequence of the $\bmath{j}_{\rm ret}\times \bmath{B}$ force, that accelerates the plasma away from the cavities, \mpo{due to a local return current, $\bmath{j}_{\rm ret}$, that is unbalanced by the homogeneously distributed CRs} \cite[see, e.g.,][]{niemiec_2008}. 
The ion temperature (Fig.~\ref{turb}c) is \mpo{particularly high in regions of very low plasma density, an example of which are the \jn{cavity} structures at $(x/\lambda_{\rm max},y/\lambda_{\rm max})\approx (125.5, 16-20), (129, 23)$ \okr{and} $(130.5, 20.5)$, \mpof{as well as the} filaments at $(x/\lambda_{\rm max},y/\lambda_{\rm max})\approx (127, 19-21)$ \okr{and} $(138.5, 19-21)$. Likewise, the ions are cold in filaments and islands of high plasma density, for example at $(x/\lambda_{\rm max},y/\lambda_{\rm max})\approx (134.5, 20-24)$. The electron temperature largely follows that of the ions, although small differences in the locations of the most efficient heating can be noticed (Fig.~\ref{turb}d). 
The anticorrelation between plasma density and temperature indicates that plasma heating does not result from adiabatic compression in turbulently moving plasma parcels. In fact, the extent} of heating depends on the strength of the electric fields that are associated with the magnetic turbulence.

As discussed in Section~\ref{global}, the \mpo{vorticity present in the region of strongly nonlinear turbulence induces motional electric field in all directions.} In particular, the $E_x$ field component is contributed by the terms $v_yB_z$ and $v_zB_y$. One can note, that the electric field is weak or disappears completely in \mpo{most of} the density filaments. 
\jn{This is because the magnetic-field amplitude is in most locations dominated by the $B_z$ component, whose change of sign between magnetic vortices leads to the cancellation of the motional electric fields.  The plasma heating can therefore occur in neighboring regions with weaker magnetic field, in which the electric field is present and can accelerate low-density particles.} 
Correspondingly, efficient heating arises in plasma cavities where large-amplitude electric field is present. Such regions can be clearly identified in Fig.~\ref{turb} around $(x/\lambda_{\rm max},y/\lambda_{\rm max})\approx (125.5, 19.5)$ or $(130.5, 16.5)$. 

The amount of heat acquired by electrons or ions at specific locations depends on the configuration of the electromagnetic fields and the local plasma flow. 
\mpo{Motional electric field by definition \mpof{vanishes} in the local flow frame, and so particles that by their kinetic energy are indistinguishable from the bulk will follow the plasma flow without significant energization. Additional electric field is required to heat such particles. In Fig.~\ref{epot} we display the charge density of the plasma and the corresponding electrostatic field in the same region as chosen for Fig.~\ref{turb}. The Poisson equation, $\ok{\nabla \cdot} \bmath{E}=\frac{e}{\epsilon_0}(N_\mathrm{i}-N_\mathrm{e})$, provides virtually the same field as that calculated in the local plasma rest frame, but details are better visible because the local plasma frame does not need to be resolved at small scales. To be noted from the figure is that many but not all locations with high plasma temperature coincide with regions of large-amplitude electrostatic field. Even in the plasma cavities one finds electrostatic field, and the potential energy is transferred to few particles, \mpof{leading to a} high kinetic energy. The charge separation that provides electrostatic field must be driven, and plasma heating damps the fields. A direct comparison of field strength and plasma temperature is impeded by the electrostatic field being a snapshot and the temperature representing the heating history of a plasma parcel. As electrons are far more mobile \mpof{along the magnetic field} than are ions, heat is redistributed efficiently. 
\okr{Hence} the spatial variation in electron temperature appears less dramatic than that of ions.}

\subsubsection{\okr{Plasma ion \jnr{and electron} scattering}\label{ionscatt}}

\begin{figure}
\centering
\includegraphics[width=\columnwidth]{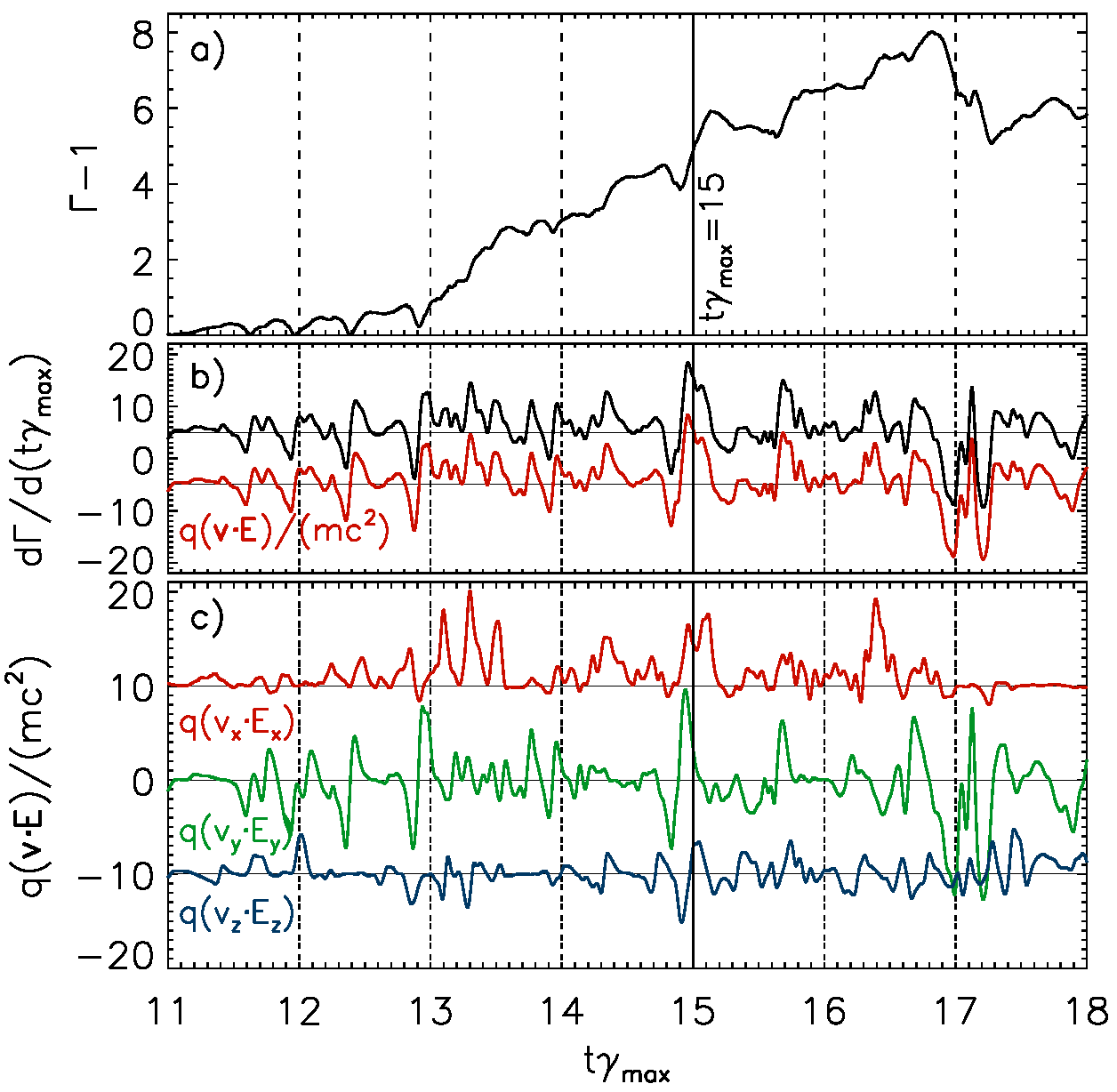}
\caption{\mpo{Panel (a): Energy evolution of a selected \jn{background} ion. Panel (b): The rate of \jn{energy} change, $d\Gamma/dt$, measured in the simulation (\textit{black} line) and the value calculated as the rate of electric work, $q(\bmath{v}\cdot\bmath{E})$, along the trajectory of the ion (\textit{red} line). The two lines are offset for clarity (+ 5 units and -5 units), and the zero level is indicated by horizontal solid lines. Panel (c): The rate of electric work decomposed into its $x, y$ \okr{and} $z$ components. All curves are smoothed.} }
\label{accelion}
\end{figure}

The \mpo{electrostatic heating applies to low-energy (thermal) electrons and ions that are tight to the bulk flow. A~few particles, \jn{that} manage to gain higher energies, are pulled out of the thermal pool and further accelerated. They sample the turbulent convective fields, undergoing stochastic scattering.
The trajectory of such an ion is shown in Fig.~\ref{turb}, 
\okr{the} details of its energization history are presented in Fig.~\ref{accelion}. A particle traveling in the turbulent field \jn{is} deflected by magnetic irregularities and scattered upon passage through regions of opposite polarity.
The motional electric field associated with the magnetic turbulence then accelerates or decelerates a particle depending on its local direction of motion, which is demonstrated in Fig.~\ref{accelion}. At time $t\gamma_{\rm max}=15$, for example, the ion is instantaneously accelerated on account of work in the $E_x$ and $E_z$ electric-field components that provide a net positive product $q(\bmath{v}\cdot\bmath{E})$ (see Fig.~\ref{accelion}b-c). Shortly before that, at $t\gamma_{\rm max}=14.8$, the particle was slightly decelerated when it moved against the \mpof{then} dominant $E_y$ electric field component. The excellent correlation between the total rate of energization and that of work in the electric field suggests that all gains and losses in ion energy measured in the simulation arise from work in the turbulent electric field. To be noted from Fig.~\ref{accelion}c is that while the work in the $E_y$ and $E_z$ fields averages to zero and would result in scattering in momentum space, the work in $E_x$ is on average positive, implying quasi-continuous gain in energy. This behaviour resembles that observed for CR ions. As noted in Section~\ref{crscatt}, CR ions tend to gain energy when moving in positive $x$-direction and to \okr{lose} energy otherwise.}

\jn{Scattering of supra-thermal electrons proceeds in a similar way as for the ions.
However, due to their small \mpof{mass compared to the ions, the electrons tend to experience} locally strong acceleration or deceleration, and abrupt energy gains and losses can occur as a particle moves through different turbulence regions. Consequently, the work of all electric-field components can contribute. 
Beside scattering in the convective electric field, some electrons can be accelerated directly in the  
$E_z$ field at the magnetic-reconnection sites that may be formed between plasma parcels. Since these sites are very few, acceleration in the reconnection structures is much less efficient than that resulting from motional electric fields.} 

\section{\okr{Discussion}} \label{discussion}

\okr{Our new realistic setup with open boundaries permits us to study the spatial structure \jnr{of a far-upstream SNR precursor region in which} the nonresonant instability \jnr{generates strong magnetic turbulence. One of the new features revealed in our study is the compression structure, reminiscent of a strong nonrelativistic supercritical shock with almost four-fold density increase.} \mpor{It arises from the difference in {the flow} speed between the incoming {quasi-}laminar and {weakly-perturbed} plasma and the plasma in the region of strong turbulence that has} already slowed down \jnr{due to backreaction effects.}}
In comparison to the energy density and inertia of the background plasma, the energy density and bulk momentum of CRs in real shock precursors of young SNRs are likely a bit smaller than they are in this simulation. The upstream plasma will always suffer some deceleration, although that might be weaker in real precursors than is observed in this study. Also, the CRs will always be pushed back, and that may be the dominant effect for realistic energy-density ratios. Hence, the level of compression will clearly depend on the acceleration efficiency of the shock. In any case, Bell's nonresonant mode is an efficient means to transfer bulk momentum from the CRs to the upstream plasma. Ignoring escape and assuming isotropy at the shock, the CR pressure in the precursor is 
$\Pi_\mathrm{CR}\approx U_\mathrm{CR}/3 $,
\mpor{where $U_\mathrm{CR}$ is the cosmic-ray energy density. In the fluid picture the gradient of the CR pressure is compensated by the gradient of the plasma momentum flux, yielding for the total change in plasma flow velocity
\begin{equation}
\delta\Pi_\mathrm{CR} =-\rho_\mathrm{up}\,V_\mathrm{up}\,\delta V \quad\Rightarrow\ \delta V=\frac{V_\mathrm{up}}{6}\,\frac{U_\mathrm{CR}}{U_\mathrm{up}}\simeq 2 v_\mathrm{A} N_\mathrm{e-fold} ,
\end{equation}
where $U_\mathrm{up}=\rho_\mathrm{up}\,V_\mathrm{up}^2 /2$ is the far-upstream bulk-flow energy density of the unshocked plasma. For the last equality we used Eqs.~10-12 of \citet{niemiec_2008} to relate the energy-density ratio to the Alfv\'en speed upstream of the shock and the number of e-foldings that is available for linear growth of Bell's instability before the plasma is captured by the shock. For standard parameters ($B_0\approx 10\ \mu$G and $n_\mathrm{up}\simeq 0.5\ \mathrm{cm}^{-3}$) we find $v_\mathrm{A}\simeq 30\ \mathrm{km/s}$, and reaching the saturation level of Bell's mode should require at least the 15 e-foldings that it needs in our simulations. The expected velocity decrement of the upstream flow is then $\delta V\approx 900\ $km/s, much larger than the sound speed, and so it is quite conceivable that also under realistic conditions a compression front develops in the shock precursor, provided Bell's instability accounts for magnetic-field amplification to saturation level. In any case, the modification to the upstream flow would be large.}

\mpor{Electromagnetic turbulence is accompanied by significant heating of background \jnr{electrons and ions. We show that this} heating does not result from adiabatic compression in density filaments formed by turbulently moving plasma parcels, as was deduced from MHD studies. Instead, the highest plasma temperatures are reached in regions of very low plasma density that harbor large-amplitude {\it electrostatic} fields. We expect that plasma heating in the precursor has impact on the properties of the shock at which CRs are accelerated, and the system will likely approach a new balance between shock acceleration, driving of the nonresonant mode \okr{and} its feedback in the precursor.}

\jnr{The scattering of CRs in space containing the turbulence amplified through the nonresonant instability is shown here to be compatible with Bohm diffusion, when the turbulence coherence scale is taken into account.}
As the precursor size scales with the mean free path for scattering,
\begin{equation}
L\simeq \frac{D}{v_\mathrm{up}}= \frac{c}{3\, v_\mathrm{up}}\,\lambda_\mathrm{mfp},
\end{equation}  
the system is about $25\,\lambda_\mathrm{mfp}$ in size, and CRs should reach the diffusive regime.
Our estimate of the spatial diffusion rate \mpo{being close to the Bohm limit is compatible with recent studies of nonresonant turbulence.}
\citet{reville08} investigated particle transport by following CR trajectories in a static snapshot of the magnetic field amplified through the nonresonant instability to nonlinear amplitude of $\delta B/B_{\parallel 0}\sim 5$, i.e., before the turbulence growth has started to slow down and saturate.
They found the diffusion rate below the Bohm limit in the pre-existing uniform field for particles with an energy considerably lower than that of CRs driving the turbulence, while $k_{\rm max}r_{\rm CRg}\gtrsim 1$ is still satisfied. \citet{reville13} ran coupled MHD-kinetic simulations that allowed for the CR filamentation instability and included the full CR dynamics, but regulated the driving to maintain an approximately constant CR current. They demonstrated sub-Bohm diffusion in the pre-existing field for CR driving the field growth, but estimated the diffusion coefficient to be comparable with the Bohm limit in the root-mean-squared magnetic-field amplitude.
Similar results were reported for full shock hybrid-kinetic simulations of strong shocks by \citet{2014ApJ...794...47C} for particles with energies close to the maximum energy, $E_{\rm max}$, propagating in strongly nonlinear turbulence generated via nonresonant instability by escaping CRs with $E\gtrsim E_{\rm max}$.

\jnr{As a final note,} the nonresonant instability, stimulating the magnetic field growth on scales small compared with the gyroradii of CRs driving it, should be accompanied by CR filamentation \citep{bell05,reville12}. 
The growth rate of the instability is proportional to the root mean square of the transverse component of the magnetic field, $\langle\delta B_\perp\rangle$, and so filamentation can \jnf{efficiently grow} only after the small-scale field has been significantly amplified. 
The scale above which filamentation dominates over the nonresonant instability can be estimated as 
$kr_{\rm CRg}=\langle\delta B_\perp/B_{\parallel 0}\rangle^2v_{\rm sh}/c$ \citep{reville12}. For our simulation the wavelength, 
$\lambda\simeq 10,000\Delta$, is comparable with the transverse size of the simulation box, and so CR
filamentation cannot be captured. \okr{However, at} late times we observe nonuniform structures in the CR \jnr{number} density, 
suggesting that the filamentation instability partially operates, although extended CR filamentary structures and the accompanying magnetic fields are not observed. Likewise, secondary MHD instabilities operating on scales larger than the CR Larmor radius cannot be described by our simulation \citep[e.g.,][]{2011MNRAS.410...39B,2010ApJ...753..6}.

\section{\okr{Summary}} \label{summary}

\mpo{We present results of a 2.5-dimensional PIC simulation of a nonresonant cosmic-ray-current-driven instability that is believed to operate at parallel shocks of young supernova remnants and provide magnetic-field amplification there.} Earlier numerical investigations of this instability used periodic boundary conditions that allow to follow only \mpo{the} temporal development of the system. They also did not account for mass conservation, \mpo{and hence did not properly describe the nonlinear backreaction of the turbulence that involves bulk deceleration of} CRs and background plasma. Our current study \mpo{should be more realistic, as it for the first time employs} open boundaries that permit plasma inflow on one side of the simulation box and outflow at the other end. This alleviates the continuity \mpo{issue and also allows studying} both the spatial and temporal evolution of the system. Our main results can be summarized as follows:
\begin{itemize}

\item We demonstrate magnetic-field amplification consistent with that found in studies with periodic simulation boxes. The evolutionary stages of the nonresonant instability, i.e., the first emergence of waves, their initial exponential growth, their attaining nonlinear wave amplitudes, their saturation \okr{and} wave cascading, that have been observed in the earlier studies only as temporal evolution, can now be mapped to specific locations in the shock precursor, and the 
relation between the spatial and the temporal development becomes visible.

\item We confirm the nonlinear backreaction of the amplified magnetic turbulence \mpo{as modification of the bulk motion of CRs and plasma. Saturation of the instability is provided by an effective convergence of the bulk velocities of CRs and ambient plasma.}

\item The magnetic field is amplified through the nonresonant instability to $\langle\delta B/B_{\parallel 0}\rangle\approx 15$, i.e., essentially the same level as observed in our earlier simulations using periodic boxes \citep{2009ApJ...706...38S}. Stronger amplification can be achieved through congestion in the region where incoming ISM plasma is decelerated \jnr{to form a compression region}.

\item The formation of \jnr{this} compression region is revealed for the first time. \jnr{The} plasma density abruptly increases \jnr{there} by a factor of $4$, and a population of ions reflected from this \jnr{shock-like} structure appears. Streaming of the reflected ions against the ambient plasma flow may cause further magnetic-field
generation via filamentation instabilities.

\item In the nonlinear phase the magnetic turbulence is accompanied by strong density and velocity fluctuations. Turbulent electric field also appears at amplitudes an order of magnitude smaller than those of the magnetic field.

\item Turbulent electric field mainly arises from magnetic-field transport, but there is also an electrostatic field component resulting from charge separation in the background plasma. On account of significant transverse plasma motions, an electric-field component in plasma drift direction with nonzero mean amplitude is present in the turbulence region. Bulk energy is transferred between CRs, the ambient plasma \okr{and} turbulence through work in this field.

\item The development of nonlinear turbulence is accompanied by significant heating of the background plasma. Ambient electrons and ions remain close to equipartition in bulk. While electrons are largely thermalized, the ions strongly deviate from a Maxwellian distribution and show supra-thermal spectral tails. \jnr{The plasma heating is not adiabatic, is most efficient in low-density regions, and results from large-amplitude {\it electrostatic} fields.}

\item Turbulent electric field inelastically scatters CRs, introducing significant anisotropy and modifying their energy distribution. Stochastic scattering off the turbulent electric field is also responsible for the generation of supra-thermal tails in the spectra of plasma ions.

\item Spatial of scattering of CRs in the turbulence generated via the nonresonant instability is compatible with Bohm diffusion, when corrected for the coherence scale of the turbulence. 

\end{itemize}

\jnr{In conclusion, our results support the notion that the nonresonant instability can produce significant magnetic-field amplification far upstream in the precursors of young SNR shocks. 
This process, when combined with additional field generation through instabilities that operate on larger scales, in close proximity to the shock front and/or at the shock itself, may be able to account for the level of magnetic-field amplification inferred from SNR observations. Similar conclusions may apply to shocks in relativistic jets and GRBs. The amplitude of the amplified field and CR scattering by the magnetic turbulence can both provide suitable conditions for particle acceleration to PeV energies at shocks in young SNRs.} \mpor{The backreaction of the magnetic-field amplification through Bell's non-resonant mode involves plasma deceleration, compression, and heating, all of which will affect the sub-shock of the thermal plasma, at which CRs are accelerated. Continuous plasma deceleration is a known effect of CRs in shock precursors \citep{1987PhR...154....1B}, but the shock-like compression feature that we observe in our simulation is a new effect whose impact on the shock structure and particle acceleration in general has yet to be explored.}

\section*{Acknowledgements}
This work has been supported by Narodowe Centrum Nauki through research projects DEC-2011/01/B/ST9/03183 (J.N.) and DEC-2013/10/E/ST9/00662 (O.K., J.N. and A.B.).
M.P. acknowledges support through grant PO 1508/1-2 of the Deutsche Forschungsgemeinschaft. The numerical experiment was possible through a 7 Mcore-hour allocation on the 2.399 PFlop Prometheus system at ACC Cyfronet AGH. Part of the numerical work was conducted on resources provided by The North-German Supercomputing Alliance (HLRN) under project bbp00003 \jnr{and also at the Pleiades facility at the NASA Advanced Supercomputing (NAS)}.

\bsp	
\label{lastpage}
\end{document}